\documentclass[journal]{IEEEtran}
\usepackage{graphicx,cite,epsfig,amssymb,amsmath}

\begin{document}
\markboth{IEEE TRANS. ON VEHICULAR TECHNOLOGY, Vol. XX,
No. Y, Feb 2016} {Ge etc.: Vehicular Communications for 5G Cooperative Small Cell Networks}
\title{\mbox{}\vspace{0.40cm}\\
\textsc{Vehicular Communications for 5G Cooperative Small Cell Networks} \vspace{0.2cm}}



\author{Xiaohu~Ge,~Hui~Cheng,~Guoqiang~Mao,~Yang~Yang,~Song~Tu

\thanks{Xiaohu~Ge, Hui~Cheng and Song~Tu  are with the School of Electronic Information and Communications, Huazhong University of Science and Technology, Wuhan 430074, Hubei, P. R. China. Emails:\{ xhge, hc\_cathy, songtu\}@hust.edu.cn.}
\thanks{Guoqiang Mao, School of Computing and Communications, University of Technology Sydney and National ICT Australia, Sydney, Australia. Email: guoqiang.mao@uts.edu.au.}
\thanks{ Yang Yang(corresponding author), Key Lab of Wireless Sensor Network and Communication, Shanghai Institute of Microsystem and Information Technology (SIMIT),
Chinese Academy of Sciences, Shanghai 200050. Email: yang.yang@wico.sh}

\thanks{The authors would like to acknowledge the support from the International Science and Technology Cooperation Program of China  under grants 2015DFG12580 and 2014DFA11640, the National Natural Science Foundation of China (NSFC) under the grants 61471180, NFSC Major International Joint Research Project under the grant 61210002, the Fundamental Research Funds for the Central Universities under the grant 2015XJGH011. This research is partially supported by the EU FP7-PEOPLE-IRSES, project acronym S2EuNet (grant no. 247083), project acronym WiNDOW (grant no. 318992) and project acronym CROWN (grant no. 610524), National international Scientific and Technological Cooperation Base of Green Communications and Networks (No. 2015B01008) and Hubei International Scientific and Technological Cooperation Base of Green Broadband Wireless Communications. G. Mao would like to acknowledge the support from ARC Discovery project DP120102030 and NSFC project 61428102. Dr Yang would like to acknowledge the support from the NSFC under grants 61231009 and 61461136003, the National Science and Technology Major Project under grant 2016ZX03001024, and the Science and Technology Commission of Shanghai Municipality (STCSM) under grant 14ZR1439700.}}

\date{\today}
\renewcommand{\baselinestretch}{1.2}
\thispagestyle{empty} \maketitle \thispagestyle{empty}
\newpage
\setcounter{page}{1}\begin{abstract}
The cooperative transmission is an effective approach for vehicular communications to improve the wireless transmission capacity and reliability in the fifth generation (5G) small cell networks. Based on distances between the vehicle and cooperative small cell BSs, the cooperative probability and the coverage probability have been derived for 5G cooperative small cell networks where small cell base stations (BSs) follow Poisson point process distributions. Furthermore, the vehicular handoff rate and the vehicular overhead ratio have been proposed to evaluate the vehicular mobility performance in 5G cooperative small cell networks. To balance the vehicular communication capacity and the vehicular handoff ratio, an optimal vehicular overhead ratio can be achieved by adjusting the cooperative threshold of 5G cooperative small cell networks.

\end{abstract}
\newpage
\IEEEpeerreviewmaketitle \vspace{-1cm}

\section{Introduction}
\label{sec1}

\IEEEPARstart{I}{n} the future fifth generation (5G) cellular networks, denser and smaller cells are expected to provide high transmission rate for users \cite{ChenZhao14,ChenZhang15,Ge16}. Different with traditional personal users, vehicles are sensitive to transmission scenarios in 5G cooperative small cell networks \cite{Maviel12,Ge14}. Moreover, due to the mobility nature of vehicles and the related high vehicular speed, the channel characteristics of the vehicular communications scenario can be significantly different from those of conventional wireless communication scenarios \cite{Cheng14,Wang07}, and make the topology of vehicular wireless networks becomes highly dynamic and prone to recurrent link intermediate \cite{ZhuBao12,Khabbaz12,ChenHao15}. In this case, cooperative transmissions are recommended as a promising solution for vehicles in 5G cooperative small cell networks \cite{LiZhangHaenggi15}. However, there still exist some problems, such as the frequent handoff and coverage problems for vehicles in 5G cooperative small cell networks \cite{YuWang14}. Therefore, it is a great challenge to investigate vehicular communications for 5G cooperative small cell networks.

   To meet the communication requirements from vehicles, some studies have been investigated for vehicular communications in cellular networks \cite{Sepulcre11,Shafiee11,Bernal-Mor15,LiXu14,LiZhangXu14}. Congestion and awareness control techniques have been investigated for cooperative vehicular communications which is based on wireless communications between vehicles and with other infrastructure nodes \cite{Sepulcre11}. To minimize the cost of transmission or alternatively transmission time in vehicular heterogeneous networks, performing verticular handoff is an appreciate choice at lower speeds, whereas it would be better to avoid vertical handoff and stay in the cellular network at higher speeds \cite{Shafiee11}. Based on a traffic model of two-tier cellular networks composed of macro cells and small cells, the impact that the user traffic dynamics, the mobility of users and the capacity constraint of the small cell backhaul on the system performance has been evaluated in \cite{Bernal-Mor15}. Accounting for the vehicular mobility and network load in cellular/802.11p heterogeneous networks, an analytical model was proposed for estimating the average achievable individual throughput and an optimal handoff threshold was derived in \cite{LiXu14}. To resolve problems resulting from limited roadside units and insufficient resources in vehicular ad hoc networks, the vehicles were configured as special vehicular small cells which have been furtherly integrated into the layered heterogeneous networks \cite{LiZhangXu14}. Considering that small cell base stations (BSs) are deployed at vehicles, a closed form outage probability was derived for evaluating the user gain in two-tier cellular networks \cite{Feteiha14}.

   Considering the coverage of small cell is smaller, the cooperative communication is widely used for small cell networks \cite{LiZhangHaenggi15,Gesbert10,Baccelli14,Tanbourgi14,Sakr14,ZhangYang12,LiHu13,Agarwal14,LiuNatarajan15}. In reference \cite{Gesbert10} multi-cell multiple-input multi-output (MIMO) cooperation concepts were examined from different perspectives, including an examination of the fundamental information-theoretic limits, a review of the coding and signal processing algorithmic developments, and consideration of scalability and system-level integration. Based on random cellular networks, a general methodology was proposed to treat problems of cooperation in cellular networks, in the case where the data exchange is allowed only between pairs of nodes \cite{Baccelli14}. Taking into account the irregular BS deployment typically encountered in practice, the signal-to-interference-plus-noise ratio (SINR) distribution with cooperation was precisely characterized in a generality-preserving form and a tractable model was furtherly proposed for analyzing noncoherent joint-transmission BS cooperations \cite{Tanbourgi14}. To mitigate the impact of the cross-tier interference in multi-tier wireless networks, a scheme was proposed for location-aware cross-tier cooperation between BSs in different tiers for downlink coordination multipoint (CoMP) transmission in two-tier cellular networks \cite{Sakr14}. Utilizing the average user throughput under CoMP and non-CoMP transmission after taking into account the downlink training overhead, each user was allowed to select transmission model between coherent CoMP and non-CoMP to avoid the extra overhead outweighing the cooperative gain in cellular networks \cite{ZhangYang12}. To fully exploit benefits of heterogeneous networks, a radio resource allocation scheme was proposed for cooperative relays where the relay nodes with in-band backhaul act as micro BSs and are able to serve users either independently or cooperatively with macro cell BSs \cite{LiHu13}. By defining the cooperation region as a function of the user quality of service (QoS) requirements and network load, a QoS aware cooperative downlink scheduling approach was proposed for cell-edge and handoff users that offers more reliability and higher effective capacity \cite{Agarwal14}. Using stochastic geometry-based heterogeneous cellular networks, the coverage probability, the average achievable rate and the energy efficient were derived for $K$-tier heterogeneous wireless networks with different cooperative sleep models for small cells \cite{LiuNatarajan15}.

   However, in all the aforementioned vehicular communication studies, only simple scenarios, such as two cooperative cells with single or multiple antennas, were considered and the underlying vehicular communications were limited to simple point-to-point wireless communications. Besides, the exact coverage probability of cooperative MIMO small cell networks with co-channel interference has not been investigated. Moreover, detailed investigation of the vehicle mobility performance used for 5G cooperative MIMO small cell networks is surprisingly rare in the open literature.
   Motivated by above gaps, in this paper we consider the scenarios of vehicular communications for vehicle-to-infrastructure (V2I) and for urban roads, we derive the vehicular handoff rate and the vehicular overhead ratio to evaluate the vehicular mobility performance in 5G cooperative MIMO small cell networks considering co-channel interference. The contributions and novelties of this paper are summarized as follows.

\begin{enumerate}
\item Based on distances between the vehicle and cooperative small cells, the cooperative probability and the coverage probability of cooperative small cell networks have been derived for vehicles equipped with multiply antennas.
\item From the proposed cooperative probability and coverage probability, the vehicle handoff rate and the vehicle overhead ratio are proposed to evaluate the vehicle mobility performance in 5G cooperative MIMO small cell networks considering co-channel interference.
\item Numerical results imply that there exists a minimum vehicle overhead ratio for 5G cooperative MIMO small cell networks considering different cooperative thresholds. This result can be used for optimizing vehicular communications in 5G cooperative MIMO small cell networks.
\end{enumerate}

The rest of this paper is organized as follows. Section II describes the system model of 5G cooperative MIMO small cell networks where small cell BSs follow Poisson point process distributions. In Section III, the cooperative probability has been derived for 5G cooperative MIMO small cell networks. Moreover, the coverage probability has been derived for 5G cooperative MIMO small cell networks considering co-channel interference from adjacent small cells in Section IV. Furthermore, in Section V the vehicular handoff rate and the vehicular overhead ratio have been proposed to evaluate the vehicular mobility performance in 5G cooperative MIMO small cell networks. Numerical results indicate that there exists a minimum vehicular overhead ratio for 5G cooperative MIMO small cell networks considering different cooperative thresholds. Finally, Section VI concludes this paper.

\section{System Model}
\label{sec2}

Fig. 1 shows the system model of 5G cooperative small cell networks which is a two-tier cellular network including macro cell BSs and small cell BSs. Macro cell BSs take charge of control information for vehicles and small cell BSs. Small cell BSs transmit the desired data to vehicles. In this case, macro cells form the control plane (C-plane) and small cells form the user plane (U-plane) in 5G cooperative small cell networks. Without loss of generality, both the control zone signalling and the L1/L2/L3 signalling, carried by physical downlink shared channel (PDSCH) and scheduled by physical downlink control channel (PDCCH) in the data zone of downlink sub-frames, are assumed to belong to the C-plane information. Only the user traffic data is carried by U-plane sub-frames in the C/U plane split architecture \cite{Song14}. Macro cells with the same radius are assumed to be regularly deployed in the infinite plane ${{{\mathbb{R}}}^2}$. Small cell BSs are assumed to be randomly deployed in the infinite plane ${{{\mathbb{R}}}^2}$. Moreover, the locations of small cell BSs follow an independent Poisson point processes ${\Phi _s}$ with the intensity ${\lambda _s}$. Every small cell BS has the same transmission power ${P_s}$. In this paper, the vehicle is assumed to be associated with the closest BS, which would suffer the least path loss during wireless transmissions. Every small cell is assumed to include only one BS and a few vehicles. Then, the cell boundary, which can be obtained through the Delaunay triangulation method by connecting the perpendicular bisector lines between each pair of small cell BSs \cite{Ferenc07}, splits the plane ${{{\mathbb{R}}}^2}$ into irregular polygons that correspond to different small cell coverage areas. Such stochastic and irregular topology forms a so-called Poisson-Voronoi tessellation (PVT) \cite{Stoyan96}. An illustration of one macro cell scenario is depicted in Fig. 1, where each small cell is denoted as ${{{{\mathfrak{E}}}_q}(q = 1,2,3, \cdots )}$. Despite of its complexity, an outstanding property of PVT random small cell networks is that the geometric characteristics of any small cell ${{{\mathfrak{E}}}_q}$ coincide with that of a typical PVT small cell ${{{\mathfrak{E}}}_1}$, according to the Palm theory \cite{Foss96}.This feature implies that the analytical results for a typical PVT small cell ${{{\mathfrak{E}}}_1}$ can be extended to the whole random small cell networks.
\begin{figure}[t!]
\vspace{0.1in}
\centerline{\includegraphics[width=3.5in]{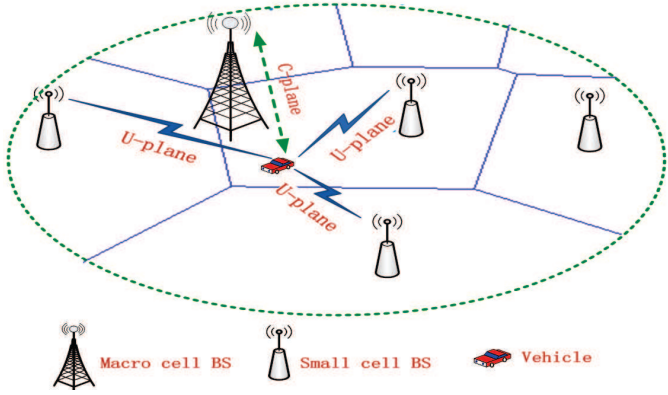}}
\caption{\small System model.}
\end{figure}

Without loss of generality, the initial location of the vehicle ${UE}_0$ located at ${{{\mathfrak{E}}}_1}$ is assumed as the origin position. The distance between the $U{E_0}$ and the closest small cell BS $B{S_1}$ is denoted as ${R_1}$. Moreover, the distance between the $U{E_0}$ and the $i - th$ closest small cell BS $B{S_i}$ is denoted as ${R_i}\left( {i = 2,3,4, \cdots } \right)$. In this paper, adjacent small cell BSs can cooperatively transmit data to a specified vehicle. Moreover, this paper is focused on downlinks of 5G cooperative small cell networks.

\section{Cooperative Probability in Small Cell Networks}
\label{sec3}

In this paper, cooperative small cell BSs is selected by distances between the vehicle  $U{E_0}$ and adjacent small cell BSs. How to evaluate the distance distribution of cooperative small cell BSs is the basis for the cooperative transmission of small cell networks.

\subsection{Distance distribution of cooperative small cell BSs}

In a homogeneous $M$-dimensionality Poisson point process with intensity $\lambda$, the probability of finding $N$ nodes in a bounded Borel space ${\cal A} \subset {{{\mathbb{R}}}^M}$ is expressed as
\[{P_r}[N\;nodes\;in\;{\cal A}] = {e^{ - \lambda {\cal A}}}\frac{{{{(\lambda {\cal A})}^N}}}{{N!}}.\tag{1}\]
For a homogeneous two-dimensionality Poisson point process with intensity $\lambda$ and ${\cal A}{\rm{ = }}\pi {r^2}$, the distance ${R_n}$ between a point and its $n-th$ closest point is governed by the generalized Gamma distribution
\[{f_{{R_n}}}(r) = {e^{ - \lambda \pi {r^2}}}\frac{{2{{(\lambda \pi {r^2})}^n}}}{{r\Gamma (n)}}\,\tag{2}\]
where $\Gamma ( \cdot )$ is the Gamma function.

{\em Corollary 1} \cite{Haenggi05}: Let ${y \in {{\mathbb{R}}}^2}$, and let ${X_i \in {{\mathbb{R}}}^2}$ be the points of a homogeneous point process of intensity $\lambda$ in ${{{\mathbb{R}}}^2}$ plane ordered according to their Euclidean distance to $y$. Then ${R_i}: = {\left\| {y - {X_i}} \right\|^2}$ has the same distribution as the one-dimensional Poisson process of intensity $\lambda \pi $, the expectation and cumulative distribution function (CDF) of  ${R_i}$ is expressed as
\[\rm \mathbb{E}[{R_i}] = i/(\lambda \pi ),\tag{3a}\]
\[{F_{{R_i}}}(r) = 1 - \frac{{{\Gamma _{ic}}(i,\lambda \pi {r^2})}}{{\Gamma (i)}},\tag{3b}\]
where ${\Gamma _{ic}}( \cdot , \cdot )$ is the incomplete Gamma function. When the differential is operated on (3b), the probability density function (PDF) of ${R_i}$ is derived by
\[{f_{{R_i}}}(r) = \frac{{2{e^{ - \lambda \pi {r^2}}}{{(\lambda \pi {r^2})}^i}}}{{r\Gamma (i)}}.\tag{4}\]

\subsection{Cooperative probability}

Since the radius of 5G small cells is usually less than 100 meters (m), the vehicle has to frequently execute the handoff operation when the high speed vehicle is only associated with one small cell BS. Even so, it is a great challenge to keep the wireless link reliability for vehicular communications in 5G small cell networks. To solve these problems, the cooperative transmission based on adjacent small cell BSs is a promising candidate. In this paper, cooperative small cell BSs is selected according to the following cooperative scheme. Considering that the radius of small cell is much less than the radius of macro cells, the wireless link is assumed be line of sight (LOS) transmission in this study. To simplify derivations, the path loss and Rayleigh fading are considered but the shadowing effect is ignored in wireless channels, as commonly done in the area \cite{Wang09,Fan15}.

\textbf{\emph{Cooperative scheme:}} When the ratio of the distance ${R_i}$ to the distance  ${R_1}$ is less than or equal to the given cooperative threshold $\rho $, the small cell BS ${BS_i}$ being ${R_i}$ apart from the vehicle ${UE_0}$ is selected for cooperative transmissions, which can be expressed as
\[\frac{{{R_i}}}{{{R_1}}} \le \rho.\tag{5} \]

Therefore, the cooperative probability of the BS ${BS_i}$ is expressed as
\[\begin{array}{l}
{P_r} (\frac{{{R_i}}}{{{R_1}}} \le \rho )\\
 = \int_0^{ + \infty } {{P_r} ({R_i} \le \rho y,{R_1} = y)} dy\\
 = \int_0^{ + \infty } {{P_r} ({R_i} \le \rho y|{R_1} = y){P_r} ({R_1} = y)} dy\\
 = \int_0^{ + \infty } {\left[ {\sum\limits_{k = i - 1}^{ + \infty } {{e^{ - {\lambda _s}{\mathop{\rm \mathfrak{D}}\nolimits} }} \cdot \frac{{{{({\lambda _s}{\mathop{\rm \mathfrak{D}}\nolimits} )}^k}}}{{k!}}} } \right]{f_{{R_1}}}} (y)dy
\end{array},\tag{6}\]
where $\mathfrak{D} = \pi {(\rho y)^2} - \pi {y^2}$ is the area between circles with different radii of ${R_1}$ and ${{\rho}R_1}$. Substitute (4) into (6), the cooperative probability of the BS ${BS_i}$ is further derived by formula (7).

\begin{figure*}[!t]
\[{P_r}(\frac{{{R_i}}}{{{R_1}}} \le \rho )
= \int_0^\infty  {\left\{ {1 - {e^{ - {\lambda _s}[\pi {{(\rho y)}^2} - \pi {y^2}]}}
\cdot \frac{{\Gamma (i - 1,{\lambda _s}[\pi {{(\rho y)}^2} - \pi {y^2}])}}{{\Gamma (i - 1)}}} \right\}}
\cdot 2{\lambda _s}\pi y{e^{ - {\lambda _s}\pi {y^2}}}dy.\tag{7}\]
\end{figure*}

 When ${k}$ small cell BSs are closed with the vehicle ${UE_0}$, the cooperative probability of $k$ adjacent small cell BSs is derived by
\[\begin{array}{l}
{\mathbb{ P} _k}= {P_r}(\frac{{{R_k}}}{{{R_1}}} \le \rho  \cap \frac{{{R_{k + 1}}}}{{{R_1}}} > \rho )\\
 = \int_0^{ + \infty } {{P_r} ({R_k} \le \rho y,{R_{k + 1}} > \rho y,{R_1} = y)} dy\\
= \int_0^{ + \infty } {{P_r} ({R_k} \le \rho y,{R_{k + 1}} > \rho y|{R_1} = y){P_r} ({R_1} = y)} dy\\
= \int_0^{ + \infty } {{e^{ - {\lambda _s}{\mathop{\rm \mathfrak{D}}\nolimits} }} \cdot \frac{{{{({\lambda _s}{\mathop{\rm \mathfrak{D}}\nolimits} )}^{k - 1}}}}{{(k - 1)!}} \cdot {f_{{R_1}}}} (y)dy\\
= \int_0^{ + \infty } {{e^{ - {\lambda _s}[\pi {{(\rho y)}^2} - \pi {y^2}]}} \cdot \frac{{{{[{\lambda _s}\pi {{(\rho y)}^2} - {\lambda _s}\pi {y^2}]}^{k - 1}}}}{{(k - 1)!}}
 2{\lambda _s}\pi y{e^{ - {\lambda _s}\pi {y^2}}}} dy\\
= \int_0^{ + \infty } {2{\lambda _s}\pi y{e^{ - {\lambda _s}\pi {{(\rho y)}^2}}} \cdot \frac{{{{[{\lambda _s}\pi {{(\rho y)}^2} - {\lambda _s}\pi {y^2}]}^{k - 1}}}}{{(k - 1)!}}} dy
\end{array}.\tag{8}\]

\subsection{Performance analysis of cooperative probability}

To validate the proposed cooperative probability, some performance analysis is simulated by numerical results in Fig. 2 and Fig. 3. The intensity of small cell BSs is configured as ${\lambda _s} = 1/(\pi  \times {50^2})$. Fig. 2 shows the impact of the cooperative threshold $\rho $ on the cooperative probability of the BS $BS_i$. When a small cell BSs is selected, the cooperative probability of the BS $BS_i$ increases with the increase of the cooperative threshold $\rho $. In this paper, the cooperative small cell BSs are ordered by the distance between the BS $BS_i$ and the vehicle $UE_0$. When the threshold $\rho $ is fixed, the cooperative probability of the BS $BS_i$ decreases with the increase of the distance between the BS $BS_i$ and the vehicle $UE_0$. When the threshold $\rho $ is larger than 3.5, the cooperative probability of the BS $B{S_i},{\rm{ }}\left( {i = 2,3,4} \right)$, approaches a saturated value.
\begin{figure}
\vspace{0.1in}
\centerline{\includegraphics[width=3.5in]{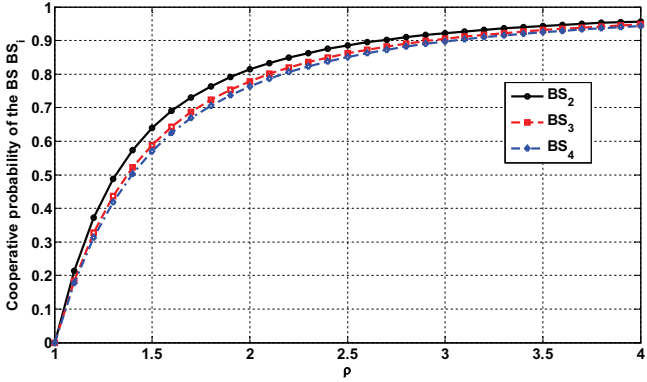}}
\caption{\small Cooperative probability with respect to the cooperative threshold  considering different cooperative small cell BSs.}
\end{figure}

Fig. 3 illustrates the impact of the number of cooperative small cell BSs and the cooperative threshold $\rho $ on the cooperative probability of small cell BSs. When the number of cooperative small cell BSs is fixed as 1, i.e., only one small cell BS is selected for cooperative transmissions, the cooperative probability monotonously decreases with the increase of the cooperative threshold $\rho $. When the number of cooperative small cell BSs is larger than 1, the cooperative probability first increases with the increase of the cooperative threshold $\rho $. When the cooperative probability achieves the maximum, the cooperative probability decreases with the increase of the cooperative threshold $\rho $. In the end, the cooperative probability approaches to a saturated value when the threshold $\rho $ is larger than 4. When the threshold is fixed, the cooperative probability decreases with the increase of the number of cooperative small cell BSs.
\begin{figure}
\vspace{0.1in}
\centerline{\includegraphics[width=3.5in]{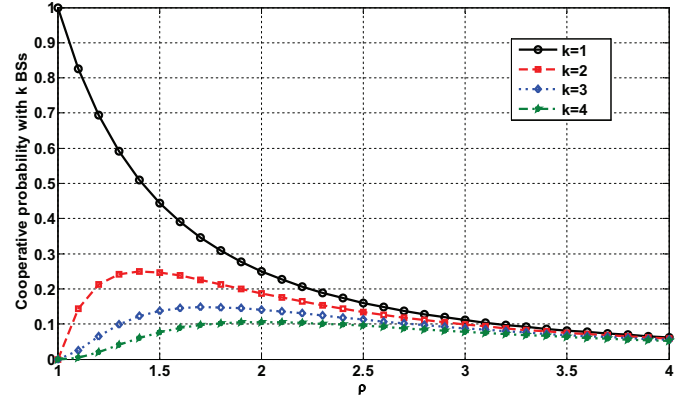}}
\caption{\small Cooperative probability with respect to the number of cooperative small cell BSs and the cooperative threshold $\rho $.}
\end{figure}

\section{Coverage Probability of Cooperative Small Cell Networks}

\subsection{Interference Model}

When the orthogonal frequency division multiplexing (OFDM) scheme is assumed to be adopted by small cell BSs to support multi-user transmission in a small cell, there is not co-channel interference generated from the intra-cell in this paper. For the vehicle $U{E_0}$, no more than one co-channel interfering vehicle is assumed to exist in each adjacent small cell. The vehicle $U{E_0}$ is interfered by downlinks of co-channel vehicles in the adjacent small cells, which is transmitted from their associated small cell BSs. The small cell BS is equipped with $n_t$ antennas and the vehicle is equipped with $n_r$ antennas. Hence, in this paper the vehicular communication is a type of MIMO communications. Without loss of generality, for the vehicle $U{E_0}$, ${\cal C} \subset {\Phi _s}$ is the cooperative small cell BSs set which can simultaneously transmit data to a given vehicle, ${\cal B} \subset {\Phi _s}\backslash {\cal C}$ is the interfering small cell BSs set. Considering the cooperative transmission from adjacent small cell BSs, the received signal at the vehicle $U{E_0}$ is expressed as
\[{\bf{y}} = \sum\limits_{i \in {\cal C}} {\frac{{\sqrt {{P_s}} }}{{R_i^{\eta /2}}}} {{\bf{H}}_{i0}}{{\bf{x}}_i} + \sum\limits_{j \in {\cal B}} {\frac{{\sqrt {{P_s}} }}{{R_j^{\eta /2}}}} {{\bf{H}}_{j0}}{{\bf{x}}_j} + {\bf{Z}},\tag{9}\]
where ${\bf{y}} \in {{\mathop{\rm \mathbb{C}}\nolimits} ^{{n_r} \times 1}}$ is the received signal vector at the vehicle $U{E_0}$, $R_i$ is the distance between the vehicle $U{E_0}$ and cooperative small cell BSs, $R_j$ is the distance between the vehicle $U{E_0}$ and interfering small cell BSs, $\eta $ is the path loss coefficient, ${{\bf{x}}_i}\in {{\mathop{\rm \mathbb{C}}\nolimits} ^{{n_t} \times 1}}$ is the desired signal vector from the cooperative transmission small cell BS $BS_i$, ${{\bf{x}}_j}\in {{\mathop{\rm \mathbb{C}}\nolimits} ^{{n_t} \times 1}}$ is the interfering signal vector from the adjacent interfering small cell  BS $BS_j$, ${\bf{Z}}\in {{\mathop{\rm \mathbb{C}}\nolimits} ^{{n_t} \times 1}}$ is the additive white Gaussian noise (AWGN) with variance ${\sigma ^2}$ in wireless channels. ${\bf{H}}_{i0}\in {{\mathop{\rm \mathbb{C}}\nolimits} ^{{n_r} \times {n_t}}}$ is the small scale fading channel matrix between the vehicle $U{E_0}$ and the cooperative small cell BS $BS_i$, ${h_{i,m,n}}{\rm{ }}\left( {m = 1,2, \cdots ,{n_r};n = 1,2, \cdots ,{n_t}} \right)$ is the element of the channel matrix ${\bf{H}}_{i0}$ and is governed by a complex Gaussian distribution, i.e., ${h_{i,m,n}}\sim{\cal C}{\cal N}(0,1)$, and its magnitude $|{h_{i,m,n}}|$ is a Rayleigh-distributed random variable \cite{Sakr14}, where ${h_{i,m,n}}$ is the channel coefficient between the $m - th$ receiving antenna at the vehicle $U{E_0}$ and the $n - th$ transmission antenna at the cooperative small cell BS $BS_i$; ${\bf{H}}_{j0}\in {{\mathop{\rm \mathbb{C}}\nolimits} ^{{n_r} \times {n_t}}}$ is the small scale fading channel matrix between the vehicle $U{E_0}$ and the interfering small cell BS $BS_j$, ${h_{j,m,n}}{\rm{ }}\left( {m = 1,2, \cdots ,{n_r};n = 1,2, \cdots ,{n_t}} \right)$ is the element of the channel matrix ${\bf{H}}_{j0}$ and is governed by a complex Gaussian distribution, i.e., ${h_{j,m,n}}\sim{\cal C}{\cal N}(0,1)$, and its magnitude $|{h_{j,m,n}}|$ is Rayleigh-distributed random variables, where ${h_{j,m,n}}$ is the channel coefficient between the $m - th$ receiving antenna at the vehicle $U{E_0}$ and the $n - th$ transmission antenna at the interfering small cell BS $BS_j$.

When $k$ cooperative small cell BSs are assumed to jointly transmit data to the vehicle $U{E_0}$, and considering the maximum ratio transmission /maximum ratio combining (MRT/MRC)\cite{Dighe03},\cite{Ge11},the SINR received by the vehicle $U{E_0}$ is derived by
\[\begin{array}{l}
SIN{R^c} = \frac{{{P_s}\sum\limits_{i \in {\cal C}}^{} {{R_i}^{ - \eta }|{{\bf{H}}_{i,0}}{|^2}} }}{{{\sigma ^2} + \sum\limits_{j \in {\cal B}} {{P_s}R_j^{ - \eta }|{{\bf{H}}_{j,0}}{|^2}} }}\\
 \quad\quad\;\;\;\;\; \approx \frac{{\frac{{{P_s}}}{{{n_t}}}\sum\limits_{i = 1}^k {{R_i}^{ - \eta }{\lambda _{\max }}({{\bf{H}}_{i,0}}{\bf{H}}_{i,0}^H)} }}{{{\sigma ^2} + \frac{{{P_s}}}{{{n_t}}}\sum\limits_{j = k + 1}^\infty  {R_j^{ - \eta }(\sum\limits_{m = 1}^{{n_r}} {\sum\limits_{n = 1}^{{n_t}} {|{h_{j,m,n}}{|^2})} } } }}\\
 \quad\quad\;\;\;\;\;\approx \frac{{\frac{{{P_s}}}{{{n_t}}}\sum\limits_{i = 1}^k {{R_i}^{ - \eta }} (\sum\limits_{m = 1}^{{n_r}} {\sum\limits_{n = 1}^{{n_t}} {|{h_{i,m,n}}{|^2})} } }}{{{\sigma ^2} + \frac{{{P_s}}}{{{n_t}}}\sum\limits_{j = k + 1}^\infty  {R_j^{ - \eta }(\sum\limits_{m = 1}^{{n_r}} {\sum\limits_{n = 1}^{{n_t}} {|{h_{j,m,n}}{|^2})} } } }}
\end{array}.\tag{10}\]
Where ${\lambda _{\max }}({\bf{H}_{i,0}}\bf{H}_{i,0}^H)$is the maximum singular value of the matrix ${\bf{H}_{i,0}}\bf{H}_{i,0}^H $. Furthermore, the interference aggregated at the vehicle $U{E_0}$ can be expressed as
\[{I_{agg}} = \frac{{{P_s}}}{{{n_t}}}\sum\limits_{j = k + 1}^\infty  {R_j^{ - \eta }(\sum\limits_{m = 1}^{{n_r}} {\sum\limits_{n = 1}^{{n_t}} {|{h_{j,m,n}}{|^2})} } } .\tag{11}\]

To simplify the derivation, let ${g_i} = \sum\limits_{m = 1}^{{n_r}} {\sum\limits_{n = 1}^{{n_t}} {|{h_{i,m,n}}{|^2}} } $ and $\Phi  \buildrel \Delta \over = \{ {R_j}|j \in {\cal B}\}$. From the distribution of ${h_{i,m,n}}$, we can derive the PDF of ${g_i}$ is ${f_{{g_i}}}(x) = \frac{{{x^{{n_t}{n_r} - 1}}}}{{\Gamma ({n_t}{n_r})}}{e^{ - x}}$. Based on the mapping theorem, $\Phi $ is an inhomogeneous Poisson point process with intensity $\lambda (r) = 2\pi {\lambda _s}r$ \cite{Haenggi09}. As a consequence, the Laplace transform of the aggregate interference at the vehicle $U{E_0}$ is derived by
\[\begin{array}{l}
{{\cal L}_{{I_{agg}}}}(s) = {\mathop{\rm \mathbb{E}}\nolimits} ({e^{ - s{I_{agg}}}})\\
= {\mathop{\rm \mathbb{E}}\nolimits} [exp( - s\sum\limits_{j = k + 1}^\infty  {\frac{{{P_s}}}{{{n_t}}}R_j^{ - \eta }(\sum\limits_{m = 1}^{{n_r}} {\sum\limits_{n = 1}^{{n_t}} {|{h_{j,m,n}}{|^2})} } } )]\\
 = {{\mathop{\rm \mathbb{E}}\nolimits} _{\Phi ,{g_j}}}\{ \mathop \Pi \limits_{j > k} [exp( - s\frac{{{P_s}}}{{{n_t}}}R_j^{ - \eta }{g_j})]\} \\
\mathop = \limits^{(a)}\exp [ - 2\pi {\lambda _s}\int_{r > {R_k}} {{{\mathop{\rm \mathbb{E}}\nolimits} _{{g_j}}}(1 - {e^{ - s\frac{{{P_s}}}{{{n_t}}}{r^{ - \eta }}{g_j}}}} )rdr]\\
 = \exp [ - 2\pi {\lambda _s}\int_{r > {R_k}} {(\int_0^\infty  {\frac{{g_{}^{{n_t}{n_r} - 1}}}{{\Gamma ({n_t}{n_r})}}{e^{ - g}}} (1 - {e^{ - s\frac{{{P_s}}}{{{n_t}}}{r^{ - \eta }}g}}} )dg)rdr]\\
 = \exp [\frac{{ - 2\pi {\lambda _s}}}{{\Gamma ({n_t}{n_r})}}\int_{r > {R_k}} {(1 - \int_0^\infty  {g_{}^{{n_t}{n_r} - 1}} {e^{ - (s\frac{{{P_s}}}{{{n_t}}}{r^{ - \eta }} + 1)g}}} dg)rdr]\\
 = \exp [ - 2\pi {\lambda _s}\int_{r > {R_k}} {(1 - \frac{1}{{{{(1 + s\frac{{{P_s}}}{{{n_t}}}{r^{ - \eta }})}^{{n_t}{n_r}}}}})} rdr]
\end{array},\tag{12}\]
where ${\mathop{\rm \mathbb{E}}\nolimits} \left(  \cdot  \right)$ is the expectation operation, (a) is due to the probability generating functional for a PPP.

\subsection{Coverage probability}

For cooperative transmissions of small cell BSs, the coverage of cooperative small cell BSs can be extended from every coverage of cooperative small cell BS. The extended coverage of cooperative small cell BSs can provide for a better reliable link service for vehicular communications. When the outage threshold is configured as $\varepsilon$ for vehicle links, the coverage probability of $k$ cooperative small cell BSs is expressed as
\[\begin{array}{l}
{\mathop{\rm \mathbb{P}}\nolimits} _c^k = {P_r}(SIN{R^c} > \varepsilon )\\
\;\;\;\;\;={P_r}\left( {\frac{{\frac{{{P_s}}}{{{n_t}}}\sum\limits_{i = 1}^k {{R_i}^{ - \eta }} {g_i}}}{{{\sigma ^2} + {I_{agg}}}} > \varepsilon } \right)
\end{array}.\tag{13}\]

However, the analytical expression can not be derived for (13) when the distance ${R_i}$ is a random variable. In the most cases, cooperative transmissions are related with the limited adjacent small cell BSs in 5G cooperative small cell networks. Moreover, the number of cooperative small cell BSs is less than or equal to 5 in realistic scenarios. When the vehicle ${UE_0}$ is assumed to be located at the edge of small cells, the distance between the vehicle ${UE_0}$ and cooperative small cell BSs is approximated to be equal. Therefore, the distance between the vehicle ${UE_0}$ and cooperative small cell BSs is configured as $D$ in the following derivations. To simplify the derivation, the transmission power of small cell BS is normalized as 1 and the noise is ignored in wireless channels considering that the noise power is obviously less than the desired signal power and the interference power \cite{Nigam14}. As a consequence, the coverage probability of $k$ cooperative small cell BSs is further derived by
\[\begin{array}{l}
{\mathop{\rm \mathbb{P}}\nolimits} _c^k = {P_r}(\sum\limits_{i = 1}^k {R_i^{ - \eta }{g_i}}  > \varepsilon {I_{agg}})\\
\;\;\;\;\,\mathop  = \limits^{(a)} {P_r}(D_{}^{ - \eta }\sum\limits_{i = 1}^k {{g_i}}  > \varepsilon {I_{agg}})\\
\;\;\;\;\,\mathop  = \limits^{(b)} {{\mathop{\rm \mathbb{E}}\nolimits} _D}\{ {{\mathop{\rm \mathbb{E}}\nolimits} _{{I_{agg}}}}[\sum\limits_{n = 0}^{k{n_t}{n_r} - 1} {\frac{{{{(\varepsilon {D^\eta })}^n}}}{{n!}}I_{agg}^n{e^{ - \varepsilon {D^\eta }{I_{agg}}}}} ]\} \\
\;\;\;\;\,\mathop  = \limits^{(c)} {{\mathop{\rm \mathbb{E}}\nolimits} _D}[\sum\limits_{n = 0}^{k{n_t}{n_r} - 1} {\frac{{{{( - \varepsilon {D^\eta })}^n}}}{{n!}}} {\cal L}_{{I_{agg}}}^{(n)}(\varepsilon {D^\eta })]
\end{array},\tag{14}\]
where the condition (a) is the assumption that the distance between the vehicle ${UE_0}$ and the cooperative small cell BSs is equal, the condition (b) is based on the CDF of Gamma distribution, the condition (c) is the Laplace transform property ${{\mathop{\rm \mathbb{E}}\nolimits} _{{I_{agg}}}}[I_{agg}^n{e^{ - s{I_{agg}}}}] = {( - 1)^n}\frac{{{d^n}}}{{d{s^n}}}{{\cal L}_{{I_{agg}}}}(s)$. Let ${x_n} = \frac{{{{( - \varepsilon {D^\eta })}^n}}}{{n!}}{\cal L}_{{I_{agg}}}^{(n)}(\varepsilon {D^\eta })$, (14) is simply expressed by
\[{\mathop{\rm \mathbb{P}}\nolimits} _c^k = {{\mathop{\rm \mathbb{E}}\nolimits} _D}\left[ {\sum\limits_{n = 0}^{k{n_t}{n_r} - 1} {{x_n}} } \right] = {{\mathop{\rm \mathbb{E}}\nolimits} _D}\left[ {{x_0} + \sum\limits_{n = 1}^{k{n_t}{n_r} - 1} {{x_n}} } \right].\tag{15}\]

 Based on (12), the differentiation of the aggregate interference at the vehicle ${UE_0}$ is derived by formula (16).
\begin{figure*}[!t]
 \[\begin{array}{l}
{\cal L}_{{I_{agg}}}^{\left( 1 \right)}\left( s \right) = \frac{d}{{ds}}{L_{{I_{agg}}}}\left( s \right) = \frac{d}{{ds}}\exp \left( { - \pi {\lambda _s}\int_{R_k^2}^\infty  {\left( {1 - \frac{1}{{{{(s{u^{ - \eta /2}} + 1)}^{{n_t}{n_r}}}}}} \right)du} } \right)\\
\quad\quad\;\;\;\;\;\; = \exp \left( { - \pi {\lambda _s}\int_{R_k^2}^\infty  {\left( {1 - \frac{1}{{{{(s{u^{ - \eta /2}} + 1)}^{{n_t}{n_r}}}}}} \right)du} } \right) \cdot \left( { - \pi {\lambda _s}\int_{R_k^2}^\infty  {\left( {1 - \frac{1}{{{{(s{u^{ - \eta /2}} + 1)}^{{n_t}{n_r}}}}}} \right)du} } \right)_s^{\left( 1 \right)}\\
\quad\quad\;\;\;\;\;\; = \exp \left( { - \pi {\lambda _s}\int_{R_k^2}^\infty  {\left( {1 - \frac{1}{{{{(s{u^{ - \eta /2}} + 1)}^{{n_t}{n_r}}}}}} \right)du} } \right) \cdot \left( { - \pi {\lambda _s}\int_{R_k^2}^\infty  {\frac{{n{}_t{n_r}{u^{ - \eta /2}}}}{{{{\left( {s{u^{ - \eta /2}} + 1} \right)}^{{n_t}{n_r} + 1}}}}du} } \right)\\
\quad\quad\;\;\;\;\;\; = \underbrace {\left( { - \pi {\lambda _s}\int_{R_k^2}^\infty  {\frac{{n{}_t{n_r}{u^{ - \eta /2}}}}{{{{\left( {s{u^{ - \eta /2}} + 1} \right)}^{{n_t}{n_r} + 1}}}}du} } \right)}_{\omega \left( s \right)}\quad  \times \quad \underbrace {{{\cal L}_I}\left( s \right)}_{v\left( s \right)}
\end{array}.\tag{16}\]
\end{figure*}

Furthermore, the $n - th$ order derivative of (12) is derived by formula (17).
\begin{figure*}[!t]
\[\begin{array}{l}
{{\cal L}}_{{I_{agg}}}^{\left( n \right)}\left( s \right) = {\left( {{{\cal L}}_I^{\left( 1 \right)}\left( s \right)} \right)^{\left( {n - 1} \right)}} = \sum\limits_{i = 0}^{n - 1} {C_{n - 1}^i{\omega ^{\left( {n - 1 - i} \right)}}{v^{\left( i \right)}}} \\
\quad\quad\;\;\;\;\;\; = \sum\limits_{i = 0}^{n - 1} {C_{n - 1}^i{\omega ^{\left( {n - 1 - i} \right)}} \times {{\cal L}}_I^{\left( i \right)}\left( s \right)} \\
\quad\quad\;\;\;\;\;\; = \pi {\lambda _s}{n_t}{n_r}\sum\limits_{i = 0}^{n - 1} {C_{n - 1}^i{{\left( { - \int_{R_k^2}^\infty  {\left( {{u^{ - {\eta  \mathord{\left/
 {\vphantom {\eta  2}} \right.
 \kern-\nulldelimiterspace} 2}}}{{\left( {1 + s{u^{ - {\eta  \mathord{\left/
 {\vphantom {\eta  2}} \right.
 \kern-\nulldelimiterspace} 2}}}} \right)}^{ - {n_t}{n_r} - 1}}} \right)du} } \right)}^{\left( {n - 1 - i} \right)}} \times {{\cal L}}_I^{\left( i \right)}\left( s \right)} \\
\quad\quad\;\;\;\;\;\; = \pi {\lambda _s}{n_t}{n_r}\sum\limits_{i = 0}^{n - 1} {C_{n - 1}^i\left( -{\int_{R_k^2}^\infty  {\left( {{u^{ - {\eta  \mathord{\left/
 {\vphantom {\eta  2}} \right.
 \kern-\nulldelimiterspace} 2}}}{{\left( { - 1} \right)}^{n - i - 1}}\frac{{\left( {n - i - 1 + {n_t}{n_r}} \right)!}}{{({n_t}{n_r})!}}} \right.} } \right.} \\
\quad\quad\;\;\;\;\;\;\;\;\; \left. {\left. { \times {{\left( {1 + s{u^{ - {\eta  \mathord{\left/
 {\vphantom {\eta  2}} \right.
 \kern-\nulldelimiterspace} 2}}}} \right)}^{ - {n_t}{n_r} - 1 - n + i + 1}}{{\left( {{u^{ - {\eta  \mathord{\left/
 {\vphantom {\eta  2}} \right.
 \kern-\nulldelimiterspace} 2}}}} \right)}^{n - i - 1}}} \right)du} \right) \times {{\cal L}}_I^{\left( i \right)}\left( s \right)\\
\quad\quad\;\;\;\;\;\; = \pi {\lambda _s}\sum\limits_{i = 0}^{n - 1} {C_{n - 1}^i{{\left( { - 1} \right)}^{n - i}}\frac{{\left( {n - i - 1 + {n_t}{n_r}} \right)!}}{{({n_t}{n_r} - 1)!}}\left( {\int_{R_k^2}^\infty  {\left( {\frac{{{{\left( {{u^{ - {\eta  \mathord{\left/
 {\vphantom {\eta  2}} \right.
 \kern-\nulldelimiterspace} 2}}}} \right)}^{n - i}}}}{{{{\left( {1 + s{u^{ - {\eta  \mathord{\left/
 {\vphantom {\eta  2}} \right.
 \kern-\nulldelimiterspace} 2}}}} \right)}^{n - i + {n_t}{n_r}}}}}} \right)du} } \right) \times {{\cal L}}_I^{\left( i \right)}\left( s \right)}
\end{array}.\tag{17}\]
 \end{figure*}

Let ${v^{ - {\eta  \mathord{\left/
 {\vphantom {\eta  2}} \right.
 \kern-\nulldelimiterspace} 2}}} = s{u^{ - {\eta  \mathord{\left/
 {\vphantom {\eta  2}} \right.
 \kern-\nulldelimiterspace} 2}}}$, then $du = {s^{\eta /2}}dv$. (17) can be further derived by
 \[\begin{array}{l}
{{\cal L}}_{{I_{agg}}}^{\left( n \right)}\left( s \right) = \pi {\lambda _s}\sum\limits_{i = 0}^{n - 1} {C_{n - 1}^i \cdot {{\left( { - 1} \right)}^{n - i}} \cdot \frac{{\left( {n - i - 1 + {n_t}{n_r}} \right)!}}{{({n_t}{n_r} - 1)!}}{s^{{2 \mathord{\left/
 {\vphantom {2 \eta }} \right.
 \kern-\nulldelimiterspace} \eta } - n + i}}} \\
\quad\quad\;\;\;\;\;\;\;\;\;\; \times \int\limits_{{\varepsilon ^{ - {2 \mathord{\left/
 {\vphantom {2 \eta }} \right.
 \kern-\nulldelimiterspace} \eta }}}}^\infty  {\frac{{{{\left( {{v^{ - {\eta  \mathord{\left/
 {\vphantom {\eta  2}} \right.
 \kern-\nulldelimiterspace} 2}}}} \right)}^{n - i}}}}{{{{\left( {1 + {v^{ - {\eta  \mathord{\left/
 {\vphantom {\eta  2}} \right.
 \kern-\nulldelimiterspace} 2}}}} \right)}^{n - i + {n_t}{n_r}}}}}dv}  \times {{\cal L}}_{{I_{agg}}}^{\left( i \right)}\left( s \right)
\end{array}.\tag{18}\]

 Substitute $s = \varepsilon {D^\eta }$ into (12), the following expression is derived as
 \[{x_0} = {{\cal L}_{{I_{agg}}}}\left( s \right) = \exp \left( { - \pi {\lambda _s}{k_0}D_{}^2} \right),\tag{19a}\]
 with
 \[{k_0} = {\varepsilon ^{{2 \mathord{\left/
 {\vphantom {2 \eta }} \right.
 \kern-\nulldelimiterspace} \eta }}}\int_{{\varepsilon ^{{{ - 2} \mathord{\left/
 {\vphantom {{ - 2} \eta }} \right.
 \kern-\nulldelimiterspace} \eta }}}}^\infty  {\left( {1 - \frac{1}{{{{(1 + {v^{{{ - \eta } \mathord{\left/
 {\vphantom {{ - \eta } 2}} \right.
 \kern-\nulldelimiterspace} 2}}})}^{{n_t}{n_r}}}}}} \right)dv}.\tag{19b} \]
 When $n \ge 1$, the following expression is derived by formula (20a-b).

 \begin{figure*}[!t]
 \[\begin{array}{l}
{x_n} = \frac{{{s^n}}}{{n!}}{\left( { - 1} \right)^n}{\cal L}_{{I_{agg}}}^{\left( n \right)}\left( s \right)\\
\quad\; = \pi {\lambda _s}\sum\limits_{i = 0}^{n - 1} {C_{n - 1}^i{{\left( { - 1} \right)}^i}\frac{{\left( {n - i - 1 + {n_t}{n_r}} \right)!}}{{n!({n_t}{n_r} - 1)!}}{s^{\frac{2}{\eta } + i}}\left( {\int_{{\varepsilon ^{ - 2/\eta }}}^\infty  {\frac{{{{\left( {{\nu ^{ - \frac{\eta }{2}}}} \right)}^{n - i}}}}{{{{\left( {1{\rm{ + }}{\nu ^{ - \frac{\eta }{2}}}} \right)}^{n - i + {n_t}{n_r}}}}}dv} } \right){\cal L}_{{I_{agg}}}^{\left( i \right)}\left( s \right)} \\
\quad\; = \pi {\lambda _s}{s^{\frac{2}{\eta }}}\sum\limits_{i = 0}^{n - 1} {C_{n - 1}^i\frac{{i!\left( {n - i - 1 + {n_t}{n_r}} \right)!}}{{n!({n_t}{n_r} - 1)!}}\left( {\int_{{\varepsilon ^{ - 2/\eta }}}^\infty  {\frac{{{{\left( {{\nu ^{ - \frac{\eta }{2}}}} \right)}^{n - i}}}}{{{{\left( {1{\rm{ + }}{\nu ^{ - \frac{\eta }{2}}}} \right)}^{n - i + {n_t}{n_r}}}}}dv} } \right){{\left( { - 1} \right)}^i}\frac{{{s^i}}}{{i!}}{\cal L}_{{I_{agg}}}^{\left( i \right)}\left( s \right)} \\
\quad\; = \pi {\lambda _s}{s^{\frac{2}{\eta }}}\sum\limits_{i = 0}^{n - 1} {\frac{{\left( {n - 1} \right)!}}{{i!\left( {n - i - 1} \right)!}}\frac{{i!\left( {n - i - 1 + {n_t}{n_r}} \right)!}}{{n!({n_t}{n_r} - 1)!}}\left( {\int_{{\varepsilon ^{ - 2/\eta }}}^\infty  {\frac{{{{\left( {{\nu ^{ - \frac{\eta }{2}}}} \right)}^{n - i}}}}{{{{\left( {1{\rm{ + }}{\nu ^{ - \frac{\eta }{2}}}} \right)}^{n - i + {n_t}{n_r}}}}}dv} } \right){x_i}} \\
\;\; \mathop  = \limits^{s = \varepsilon {D^\eta }} \pi {\lambda _s}{D^2}\sum\limits_{i = 0}^{n - 1} {\frac{{{n_t}{n_r}C_{n - 1 - i + {n_t}{n_r}}^{{n_t}{n_r}}}}{n}{\varepsilon ^{2/\eta }}\left( {\int_{{\varepsilon ^{ - \frac{2}{\eta }}}}^\infty  {\frac{{{{\left( {{\nu ^{ - \frac{\eta }{2}}}} \right)}^{n - i}}}}{{{{\left( {1{\rm{ + }}{\nu ^{ - \frac{\eta }{2}}}} \right)}^{n - i + {n_t}{n_r}}}}}dv} } \right){x_i}} \\
\quad\; = \pi {\lambda _s}D_{}^2\sum\limits_{i = 0}^{n - 1} {\frac{{{n_t}{n_r}C_{n - 1 - i + {n_t}{n_r}}^{{n_t}{n_r}}}}{n}{k_{n - i}}{x_i}}
\end{array},\tag{20a}\]
with
\[{k_i} = {\varepsilon ^{2/\eta }}\int\limits_{{\varepsilon ^{ - {2 \mathord{\left/
 {\vphantom {2 \eta }} \right.
 \kern-\nulldelimiterspace} \eta }}}}^\infty  {\frac{1}{{{{\left( {1 + {v^{{\eta  \mathord{\left/
 {\vphantom {\eta  2}} \right.
 \kern-\nulldelimiterspace} 2}}}} \right)}^i}{{\left( {1 + {v^{ - {\eta  \mathord{\left/
 {\vphantom {\eta  2}} \right.
 \kern-\nulldelimiterspace} 2}}}} \right)}^{{n_t}{n_r}}}}}dv} \quad i \ge 1.\tag{20b}\]
 \end{figure*}

Based on (20a), a linear recurrence relation of ${x_n}$ is derived for the explicit expression of the coverage probability via linear algebra. Let
\[{\bf x_{k{n_t}{n_r}}} = {[{x_1},{x_2}, \cdots {x_{k{n_t}{n_r}}}]^T},\tag{21a}\]
\[\begin{array}{l} {\bf y_{k{n_t}{n_r}}} = {[{y_1},{y_2}, \cdots {y_{k{n_t}{n_r}}}]^T} \\
= {[{n_t}{n_r}{k_1},\frac{{{n_t}{n_r}(1 + {n_t}{n_r})}}{2}{k_2}, \cdots \frac{{{n_t}{n_r}C_{k{n_t}{n_r} - 1 + {n_t}{n_r}}^{{n_t}{n_r}}}}{{k{n_t}{n_r}}}{k_{k{n_t}{n_r}}}]^T}
\end{array},\tag{21b}\]

then (20) can be represented in a matrix form as formula (22a-c).
\begin{figure*}[!t]
\[{\bf x_{k{n_t}{n_r}}} = a{x_0}{\bf y_{k{n_t}{n_r}}} + a{\bf G_{k{n_t}{n_r}}}{\bf x_{k{n_t}{n_r}}},\tag{22a}\]
\[a = \pi {\lambda _s}{D^2},\tag{22b}\]
\[{\bf G_{k{n_t}{n_r}}} = \left[ {\begin{array}{*{20}{c}}
0&{}&{}&{}&{}\\
{\frac{1}{2}{n_t}{n_r}{k_1}}&0&{}&{}&{}\\
{\frac{{{n_t}{n_r}(1 + {n_t}{n_r})}}{3}{k_2}}&{\frac{{{n_t}{n_r}}}{3}{k_1}}&0&{}&{}\\
 \vdots &{}&{}&0&{}\\
{\frac{{{n_t}{n_r}C_{k{n_t}{n_r} + {n_t}{n_r} - 2}^{{n_t}{n_r}}}}{{k{n_t}{n_r}}}{k_{k{n_t}{n_r} - 1}}}&{\frac{{{n_t}{n_r}C_{k{n_t}{n_r} + {n_t}{n_r} - 3}^{{n_t}{n_r}}}}{{k{n_t}{n_r}}}{k_{k{n_t}{n_r} - 2}}}& \cdots &{\frac{{{n_t}{n_r}}}{{k{n_t}{n_r}}}{k_1}}&0
\end{array}} \right].\tag{22c}\]
\end{figure*}
Since ${\bf G_{k{n_t}{n_r}}}$ is a strictly lower triangular matrix, we have $ {\bf G_{k{n_t}{n_r}}}^n = 0,n \ge k{n_t}{n_r}$. According to this property, after iterating, ${\bf x_{k{n_t}{n_r}}}$ can be rewritten as
\[\begin{array}{l}
{\bf x_{k{n_t}{n_r}}} = a{x_0}{\bf y_{k{n_t}{n_r}}} + a{\bf G_{k{n_t}{n_r}}}{\bf x_{k{n_t}{n_r}}}\\
\quad\quad\;\;\; = a{x_0}{\bf y_{k{n_t}{n_r}}} + a{\bf G_{k{n_t}{n_r}}}(a{x_0}{\bf y_{k{n_t}{n_r}}} + a{\bf G_{k{n_t}{n_r}}}{\bf x_{k{n_t}{n_r}}})\\
\quad\quad\;\;\; = a{x_0}{\bf y_{k{n_t}{n_r}}} + {a^2}{x_0}{\bf G_{k{n_t}{n_r}}}{\bf y_{k{n_t}{n_r}}} + {a^2} {\bf {G_{k{n_t}{n_r}}}}^2{\bf x_{k{n_t}{n_r}}}\\
\quad\quad\;\;\; =  \cdots \\
\quad\quad\;\;\; = \sum\limits_{n = 1}^{k{n_t}{n_r}} {{a^n}} {x_0} {\bf G_{k{n_t}{n_r}}}^{n - 1}{\bf y_{k{n_t}{n_r}}}
\end{array}.\tag{23}\]
Similarly, denote ${\bf x_{k{n_t}{n_r} - 1}} = {[{x_1},{x_2}, \cdots {x_{k{n_t}{n_r} - 1}}]^T}$, then
\[{\bf x_{k{n_t}{n_r} - 1}} = \sum\limits_{n = 1}^{k{n_t}{n_r} - 1} {{a^n}} {x_0} {\bf G_{k{n_t}{n_r} - 1}}^{n - 1}{\bf y_{k{n_t}{n_r} - 1}}.\tag{24}\]
Define $sum({\bf x_{k{n_t}{n_r} - 1}}) = \sum\limits_{n = 1}^{k{n_t}{n_r} - 1} {{x_n}} $, which is the sum of elements in the vector  ${\bf x_{k{n_t}{n_r} - 1}}$. In the end, the coverage probability of $k$ cooperative small cell BSs can be expressed as an explicit form as (25).

\begin{figure*}[!t]
\[\begin{array}{l}
{\mathop{\rm \mathbb{P}}\nolimits} _c^k = {{\mathop{\rm \mathbb{E}}\nolimits} _D}\left[ {{x_0} + \sum\limits_{n = 1}^{k{n_t}{n_r} - 1} {{x_n}} } \right]\\
\;\;\;\;\; = \int_0^\infty  {{f_D}(D)} ({x_0} + \sum\limits_{n = 1}^{k{n_t}{n_r} - 1} {{x_n}} )dD\\
\;\;\;\;\; = \int_0^\infty  {{f_D}{{(D)}^{}}} (\exp \left( { - \pi {\lambda _s}D_{}^2{\varepsilon ^{{2 \mathord{\left/
 {\vphantom {2 \eta }} \right.
 \kern-\nulldelimiterspace} \eta }}}\int_{{\varepsilon ^{{{ - 2} \mathord{\left/
 {\vphantom {{ - 2} \eta }} \right.
 \kern-\nulldelimiterspace} \eta }}}}^\infty  {\left( {1 - \frac{1}{{{{(1 + {v^{{{ - \eta } \mathord{\left/
 {\vphantom {{ - \eta } 2}} \right.
 \kern-\nulldelimiterspace} 2}}})}^{{n_t}{n_r}}}}}} \right)dv} } \right) + sum({\bf x_{k{n_t}{n_r}-1}}))dD\\
\end{array}.\tag{25}\]
\end{figure*}

Without loss of generality, three small cell cooperative BSs scenario is considered as follows. In this case, the distance $D$ between the vehicle $UE_0$ and cooperative small cell BSs is given by \cite{Nigam14}
\[{f_D}(D) = 2{(\lambda \pi )^2}{D^3}{e^{ - \lambda \pi {D^2}}}.\tag{26}\]
Substitute (26) into (25), the coverage probability of three cooperative small cell BSs is simply derived by (27).

\begin{figure*}[!t]
\[\begin{array}{l}
\mathbb{P}_c^3 = \int_0^\infty  {2{{(\lambda \pi )}^2}{D^3}{e^{ - \lambda \pi {D^2}}}} \left( {\exp } \right.\left( { - \pi {\lambda _s}D_{}^2{\varepsilon ^{{2 \mathord{\left/
 {\vphantom {2 \eta }} \right.
 \kern-\nulldelimiterspace} \eta }}}\int_{{\varepsilon ^{{{ - 2} \mathord{\left/
 {\vphantom {{ - 2} \eta }} \right.
 \kern-\nulldelimiterspace} \eta }}}}^\infty  {\left( {1 - \frac{1}{{{{(1 + {v^{{{ - \eta } \mathord{\left/
 {\vphantom {{ - \eta } 2}} \right.
 \kern-\nulldelimiterspace} 2}}})}^{{n_t}{n_r}}}}}} \right)dv} } \right)\\
\quad\quad\; \left. { + sum({\bf x_{k{n_t}{n_r}-1}})} \right)dD
\end{array}.\tag{27}\]
\end{figure*}

\subsection{Performance analysis of coverage probability}
Based on the proposed coverage probability of three cooperative small cell BSs in (27), some performance evaluations are numerically analyzed in Fig. 4--Fig. 7. In the following coverage probability analysis,  default parameters are configured as follows: the antenna number at the small cell BS is ${n_t} = 4$, the antenna number at the vehicle is ${n_r} = 2$, the small cell BS transmission power is normalized as 1, the path loss coefficient is $\eta  = 4$, the intensity of small cell BSs is ${\lambda _s} = 1/(\pi  \times {50^2})$, the outage threshold of vehicle links is $\varepsilon  = 0$ dB \cite{LiZhangLetaief14}, the radius of small cell is 50 m.
\begin{figure}
\vspace{0.1in}
\centerline{\includegraphics[width=3.5in,height=2in,]{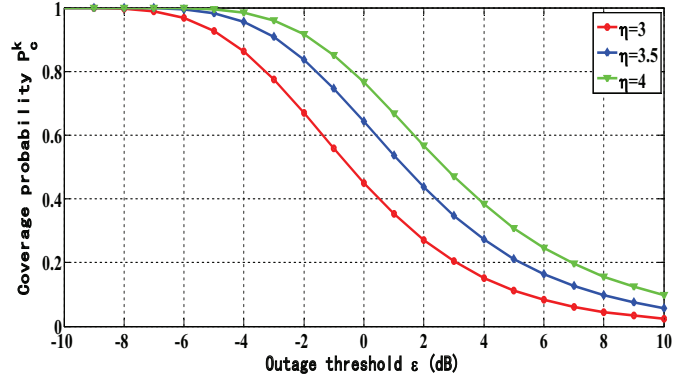}}
\caption{\small  Coverage probability with respect to the outage threshold considering different path loss coefficients.}
\end{figure}

\begin{figure}
\vspace{0.1in}
\centerline{\includegraphics[width=3.5in]{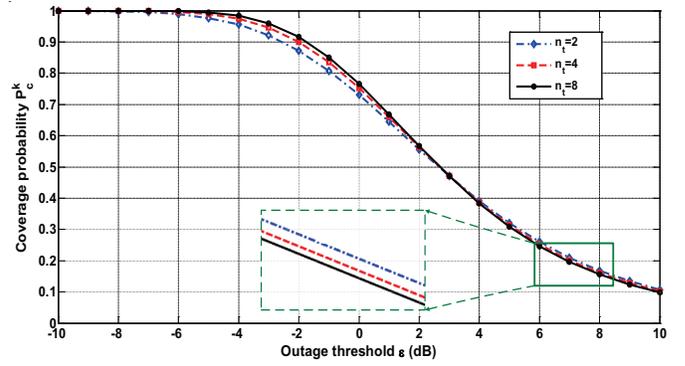}}
\caption{\small  Coverage probability with respect to the number of transmission antennas at cooperative small cell BSs considering different outage thresholds.}
\end{figure}
Fig. 4 illustrates the impact of the outage threshold and the path loss coefficient on the cooperative probability. When the path loss coefficient is fixed, the coverage probability decreases with the increase of the outage threshold. When the outage threshold is fixed, the coverage probability increases with the increase of the path loss coefficient. This result can be explained by the follows: the signal propagation attenuation is obviously increased with the increase of the propagation range when the path loss coefficient is increased. Compared with cooperative small cell BSs, the interfering small cell BSs is far away with the received vehicle. Therefore, the interference attenuation is larger than the desired signal attenuation. When the outage threshold and the transmission power of small cell BS are fixed, the coverage probability of three cooperative small cell BSs increases with the increase of the path loss coefficient.

Fig. 5 analyze the impact of the number of transmission antennas at cooperative small cell BSs on the coverage probability. When the outage threshold is less than 2.5 dB, the coverage probability increases with the increase of the number of transmission antennas at cooperative small cell BSs. When the outage threshold is larger than or equal to 2.5 dB, the coverage probability decreases with the increase of the number of transmission antennas at cooperative small cell BSs.
\begin{figure}
\vspace{0.1in}
\centerline{\includegraphics[width=3.5in]{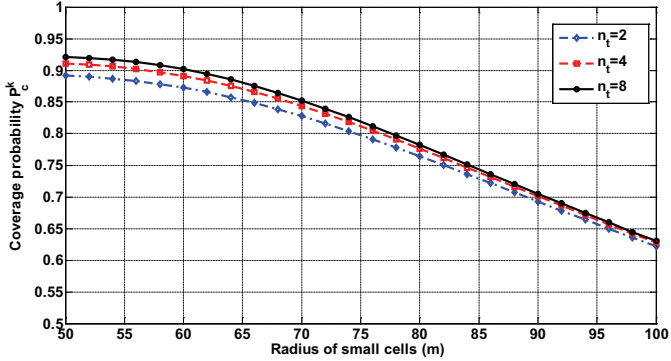}}
\caption{\small  Coverage probability with respect to the radius of small cells considering different transmission antennas at cooperative small cell BSs.}
\end{figure}
Fig. 6 depicts the impact of the radius of small cells and the number of transmission antennas at cooperative small cell BSs on the coverage probability. When the number of transmission antennas is fixed, the coverage probability decreases with the increase of the radius of small cells. When the radius of small cell is fixed, the coverage probability increases with the increase of the number of transmission antennas at cooperative small cell BSs.

Fig. 7 shows the coverage probability with and without cooperative transmission schemes in small cell networks. When the cooperative transmission scheme is adopted in small cell networks, the number of cooperative small cell BSs is configured as $k = 3$. When the cooperative communication scheme is not adopted in small cell networks, the number of cooperative small cell BSs is configured as $k = 1$. Compared with two curves in Fig. 7, the cooperative transmission scheme can improve the coverage probability in small cell networks and the gain of coverage probability with the cooperative transmission scheme achieve the maximum when the outage threshold is -1 dB.
\begin{figure}
\vspace{0.1in}
\centerline{\includegraphics[width=3.5in,height=2in,]{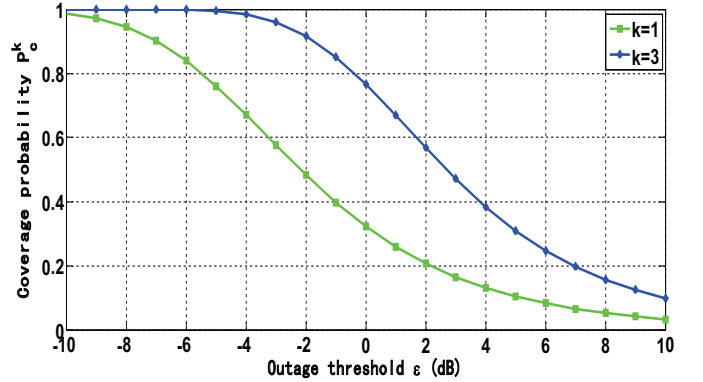}}
\caption{\small Coverage probability with and without cooperative communication schemes.}
\end{figure}

\section{Vehicle Mobility Analysis of Cooperative Small Cell Networks}
\label{sec5}
\subsection{Vehicle handoff rate of cooperative small cell networks}

When the initial position of the vehicle is assumed as the origin, the vehicle will arrive at a new position after a minimal slot $\tau $. Without loss of generality, the Gauss-Markov Mobility model is adopted for the vehicle mobility in this study. Assume that the vehicle moves with the velocity of $\varphi _{n - 1}$ and the direction of $\vartheta _{n - 1}$ at the $(n - 1)^{th}$ time instant, then the velocity and direction of the $n^{th}$ time instant are calculated by $\varphi _n = \alpha {\varphi _{n - 1}} + (1 - \alpha )\bar \varphi  + \sqrt {(1 - {\alpha ^2})} {\varphi _{{x_{n - 1}}}}$, $\vartheta _n = \alpha {\vartheta _{n - 1}} + (1 - \alpha )\bar \vartheta  + \sqrt {(1 - {\alpha ^2})} {\vartheta _{{x_{n - 1}}}}$ respectively, where $\bar \varphi $ and $\bar \vartheta $ represent the mean value of velocity and direction as $n \to \infty $, and $\varphi _{{x_{n - 1}}}$ and $\vartheta _{{x_{n - 1}}}$ are random variables governed by Gaussian distribution. To simplify the derivation, we assume that $\alpha  = 1$. In this case, the vehicle keeps the velocity $\varphi$ and the direction $\vartheta$ constant. When the distance between the vehicle $U{E_0}$ and the $i - th$ closest small cell BS $B{S_i}$ is assumed as ${R_i}$ in the last slot, the new distance between the vehicle $U{E_0}$ and the small cell BS $B{S_i}$ in the current slot is expressed as
\[R_i^{new} = \sqrt {R_i^2 + {{\left( {\varphi \tau } \right)}^2} + 2{R_i}\varphi \tau \cos \vartheta } .\tag{28} \]
 Let $\xi  = \arg {\rm{ }}\mathop {min}\limits_i R_i^{new}$, the distance between the vehicle $U{E_0}$ and the closest small cell BS $B{S_\xi }$ is denoted as $R_\xi ^{new}$ in the current slot. The vehicle passes through different cooperative small cells when the vehicle moves for a slot  $\tau $. The trigger of vehicle handoff in a slot  $\tau $ is expressed as (29).

\begin{figure*}[!t]
\[\Delta {H_\tau } = 1\left[ {\exists i \in {N^ + },(\frac{{R_i^{new}}}{{R_\xi ^{new}}} > \rho  \cap \frac{{{R_i}}}{{{R_1}}} \le \rho ) \cup (\frac{{R_i^{new}}}{{R_\xi ^{new}}} \le \rho  \cap \frac{{{R_i}}}{{{R_1}}} > \rho )} \right], \tag{29} \]
\end{figure*}

 where $1[ \cdot ]$ is the indicator function, which equals to 1 when the condition inside the bracket is satisfied and 0 otherwise. For a long time, e.g., $T = t\tau ,\;{\rm{ }}t \gg 1{\;\rm{ and }\;}t \in {N^ +}$ , the handoff number of the vehicle ${\Delta {H_T}}$ is the sum of (29) in the time $T$. Furthermore, the vehicle handoff rate in cooperative small cell networks is expressed by\[HO = \frac{{\Delta {H_T}}}{T}.\tag{30}\]

Fig. 8 analyzes the impact of the vehicular velocity and the cooperative threshold on the vehicular handoff rate. Without loss of generality, the traditional fixed two small cells cooperative scenario is compared with the proposed multi-cell cooperative scenarios in Fig. 8. When the cooperative threshold is fixed, the vehicular handoff rate increases with the increase of the vehicular velocity in both scenarios. When $\rho  = 1$, it means only one small cell transmits signals to the vehicle. In this case, the curve of multi-cell cooperative scenario is coincided with the curve of fixed two small cells cooperative scenario. when $\rho  \ne 1$, the number of cooperative small cells is changed accounting for the distance between the vehicle and the small cell BSs for the proposed multi-cell cooperative scenarios. Therefore, the vehicular handoff rate of the proposed multi-cell cooperative scenarios is obviously larger than the vehicular handoff rate of traditional fixed two small cells cooperative scenario.

\begin{figure}
\vspace{0.1in}
\centerline{\includegraphics[width=3.5in,draft=false]{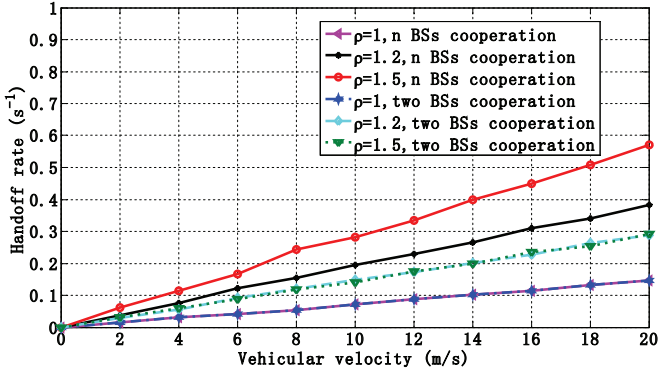}}
\caption{\small  Vehicular handoff rate with respect to the vehicular velocity and the cooperative threshold.}
\end{figure}

\subsection{Vehicle overhead ratio of cooperative small cell networks}

When the vehicle passes through different small cells, the cooperative status has to be changed. Based on small cell network architectures \cite{Robson12}, the cooperative link of small cells is defined as the X2 link, which has been used for transmitting handoff and cooperative information in cooperative small cell networks. The X2 link of cooperative small cell networks is composed of two parts, i.e., the X2-U link and the X2-C link. In general, the X2-U link is used for transmitting handoff information and the X2-C link is used for transmitting cooperative control information among small cells. Therefore, the overhead traffic of X2-C links is expressed as\[{T_{X2 - C}} = \delta  \cdot HO,\tag{31}\]
where $\delta $ is the average cooperative control data size per small cell when a handoff is triggered in small cell networks \cite{Widjaja09}.

when the total $L$ types traffic is assumed for vehicular communications, let ${\psi _l}$ and $\zeta _l^{ - 1},{\rm{ }}1 \le l \le L$, are the traffic arrive rate and the average vehicle session duration of the $type - l$ vehicle traffic. Based on the queue theory, the active probability of the $type - l$ vehicle traffic is expressed as\[{p_A}\left( l \right) = {{{\psi _l}} \mathord{\left/
 {\vphantom {{{\psi _l}} {\left( {{\psi _l} + {\zeta _l}} \right)}}} \right.
 \kern-\nulldelimiterspace} {\left( {{\psi _l} + {\zeta _l}} \right)}}.\tag{32}\]
Furthermore, the handoff rate of the $type - l$ vehicle traffic is expressed as\[H{O_l} = {p_A}\left( l \right) \cdot HO.\tag{33}\]
Let ${a_l}$ is the flow rate of $type - l$ vehicle traffic and the average handoff duration is $\chi $, the overhead of the $type - l$  vehicle traffic generated by a handoff over the X2-U link is expressed as\[{\beta _l} = {a_l} \cdot \chi. \tag{34}\]

When all handoff requests are assumed to be accepted in small cell networks, the overhead traffic of X2-U links is expressed as\[{T_{X2 - U}} = \sum\limits_{l = 1}^L {{\beta _l} \cdot H{O_l}}.\tag{35}\]

Based on (31) and (35), the total overhead traffic of X2 links in cooperative small cell networks can be expressed by\[{T_{X2}} = {T_{X2 - C}} + {T_{X2 - U}}.\tag{36}\]

Without loss of generality, a change of the cooperative status is triggered by a vehicular handoff in cooperative small cell networks. In general, the change of vehicular cooperative status conduces to some overhead traffic in cooperative small cell networks. The size of overhead traffic is expressed as\[C = k \cdot {T_{X2}},\tag{37}\]
Therefore, the expectation of the size of overhead traffic is expressed as\[\mathbb{E}[C] = \sum\limits_k {{\mathbb{P}_k} \cdot k \cdot {T_{X2}}} .\tag{38}\]

When the SINR of vehicular communication is configured as the outage threshold $\varepsilon $, the vehicular communication capacity is expressed as\[\partial  = \left[ {\sum\limits_{k = 1}^\infty  {\mathbb{P}_c^k \cdot {P_k}} } \right] \cdot {{\rm{B}}_w} \cdot \log (1 + \varepsilon ),\tag{39}\]
where ${{\rm{B}}_w}$ is the bandwidth for vehicle wireless links.

 To evaluate the cooperative communication overhead for vehicular communications, the vehicular overhead ratio in cooperative small cell networks is defined as\[\Omega  = \frac{{\mathbb{E}[C]}}{\partial }.\tag{40}\]	

 Considering the cooperative transmission in small cell networks, the gain, i.e., the vehicular communication capacity, and the cost, i.e., the vehicular overhead ratio can be evaluated by (39) and (40).

 \subsection{Performance analysis}

  Based on the proposed vehicular communication capacity and vehicular overhead ratio, some performance evaluations are numerically analyzed in Fig. 9--Fig. 11. In the following analysis, the default parameters are configured as follows: the outage threshold is $\varepsilon  = 0$ dB, the bandwidth is ${{\rm{B}}_w} = 10$ Mbps, the average cooperative control data size is $\delta  = 480$ bits, the path loss coefficient is $\eta  = 4$, the time slot is $\tau  = 15$  millisecond (ms), the radius of small cell is 50 m, the vehicular velocity is $\varphi  = 10$ m/s, the handoff duration is $\chi  = 0.05$ s. Without loss of generation, two types of traffic are considered in this paper. The $type - 1$ traffic has the following configuration parameters: ${a_1} = 12.2$ Kbps, ${\psi _1} = 1.5$ and $\zeta _1^{ - 1} = 0.03333$. The  $type - 2$ traffic has the following configuration parameters: ${a_2} = 353.8$ Kbps, ${\psi _2} = 0.5$ and $\zeta _2^{ - 1} = 0.05$ \cite{Widjaja09}.

   Fig. 9 evaluates the vehicular communication capacity with respect to the radius of small cells considering different cooperative thresholds. When the cooperative threshold is fixed, the vehicular communication capacity decreases with the increase of the radius of small cells. When the radius of small cells is fixed, the vehicular communication capacity increases with the increase of the cooperative threshold.
\begin{figure}
\vspace{0.1in}
\centerline{\includegraphics[width=3.5in]{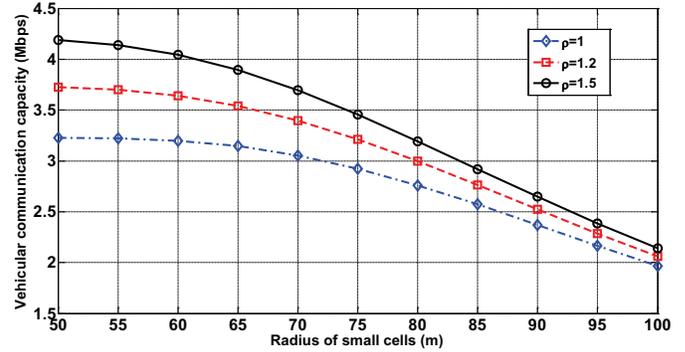}}
\caption{\small Vehicular communication capacity with respect to the radius of small cells considering different cooperative thresholds.}
\end{figure}
   Fig. 10 illustrates the vehicular overhead ratio with respect to the vehicular velocity considering different cooperative thresholds. When the cooperative threshold is fixed, the vehicular overhead ratio increases with the increase of the vehicular velocity. When the vehicular velocity is fixed, the vehicular overhead ratio increases with the increase of the cooperative threshold. The capacity gain is increased when the cooperative threshold is less than 1.5. When the cooperative threshold is larger than or equal to 1.5, the cooperation probability of $k$ small cells decreases with the increase of the cooperative threshold. This result is validated in Fig. 3. It implies that the number of cooperative small cells will be limited with the increase of the cooperative threshold in practical applications. Therefore, the capacity gain is limited with increase of the cooperative threshold.¡±
\begin{figure}
\vspace{0.1in}
\centerline{\includegraphics[width=3.5in,draft=false]{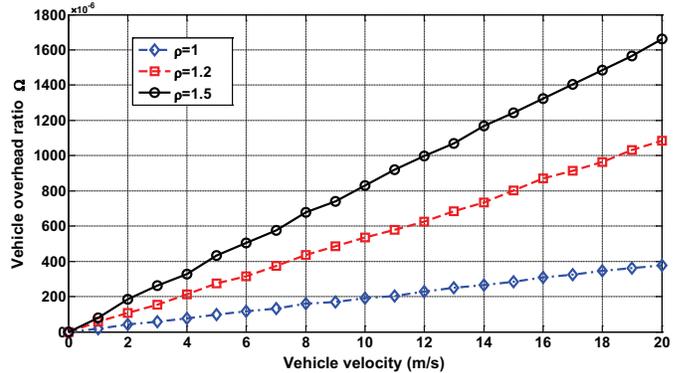}}
\caption{\small Vehicular overhead ratio with respect to the vehicular velocity considering different cooperative thresholds.}
\end{figure}
   Fig. 11 analyzes the vehicular overhead ratio with respect to the radius of small cells considering different cooperative thresholds. When the radius of small cells is fixed, the vehicular overhead ratio increases with the increase of the cooperative threshold. When the cooperative threshold is fixed, the vehicular overhead ratio first decreases with the increase of the radius of small cells. Numerical results show that there exist turning points, i.e., the radius of small cells is 75, 80 and 85 m, corresponding to the cooperative threshold 1.5, 1.2 and 1. When the radius of small cells is larger than the turning points, the vehicular overhead ratio increases with the increase of the radius of small cells. Therefore, there exist a minimal value for the vehicular overhead ratio under different cooperative thresholds. The minimal vehicle overhead ratio is $4.5 \times {10^{ - 4}}$,  $3.4 \times {10^{ - 4}}$ and  $1.3 \times {10^{ - 4}}$, corresponding to the cooperative threshold of 1, 1.2 and 1.5. This result provide a guideline for optimizing the vehicular overhead ratio in 5G cooperative small cell networks with different radii of small cells.
\begin{figure}
\vspace{0.1in}
\centerline{\includegraphics[width=3.5in]{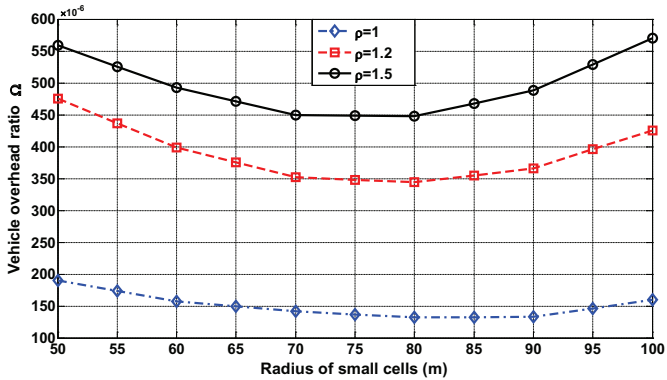}}
\caption{\small Vehicular overhead ratio with respect to the radius of small cells considering different cooperative threshold.}
\end{figure}

\section{Conclusions}
\label{sec6}

In this paper, the vehicular mobility performance is analyzed for 5G cooperative MIMO small cell networks considering co-channel interference. Based on distances between the vehicle and cooperative small cell BSs, the cooperative probability and the coverage probability have been derived for 5G cooperative small cell networks where small cell BSs follow Poisson point process distributions. Furthermore, the vehicular handoff rate and the vehicular overhead ratio have been proposed to evaluate the vehicular mobility performance in 5G cooperative MIMO small cell networks. Numerical results indicate that the cooperative transmission scheme increases the vehicular communication capacity and the vehicular handoff rate in 5G cooperative MIMO small cell networks. Therefore, there exist a tradeoff between the vehicular communication capacity and the vehicular handoff ratio. By evaluating the vehicular overhead ratio, numerical results show that there exists a minimum vehicular overhead ratio for 5G cooperative MIMO small cell networks considering different cooperative thresholds. These results provide a guideline for optimizing vehicular communications in 5G cooperative MIMO small cell networks.


\begin{IEEEbiography}[{\includegraphics[width=1in,height=1.25in,clip,keepaspectratio]{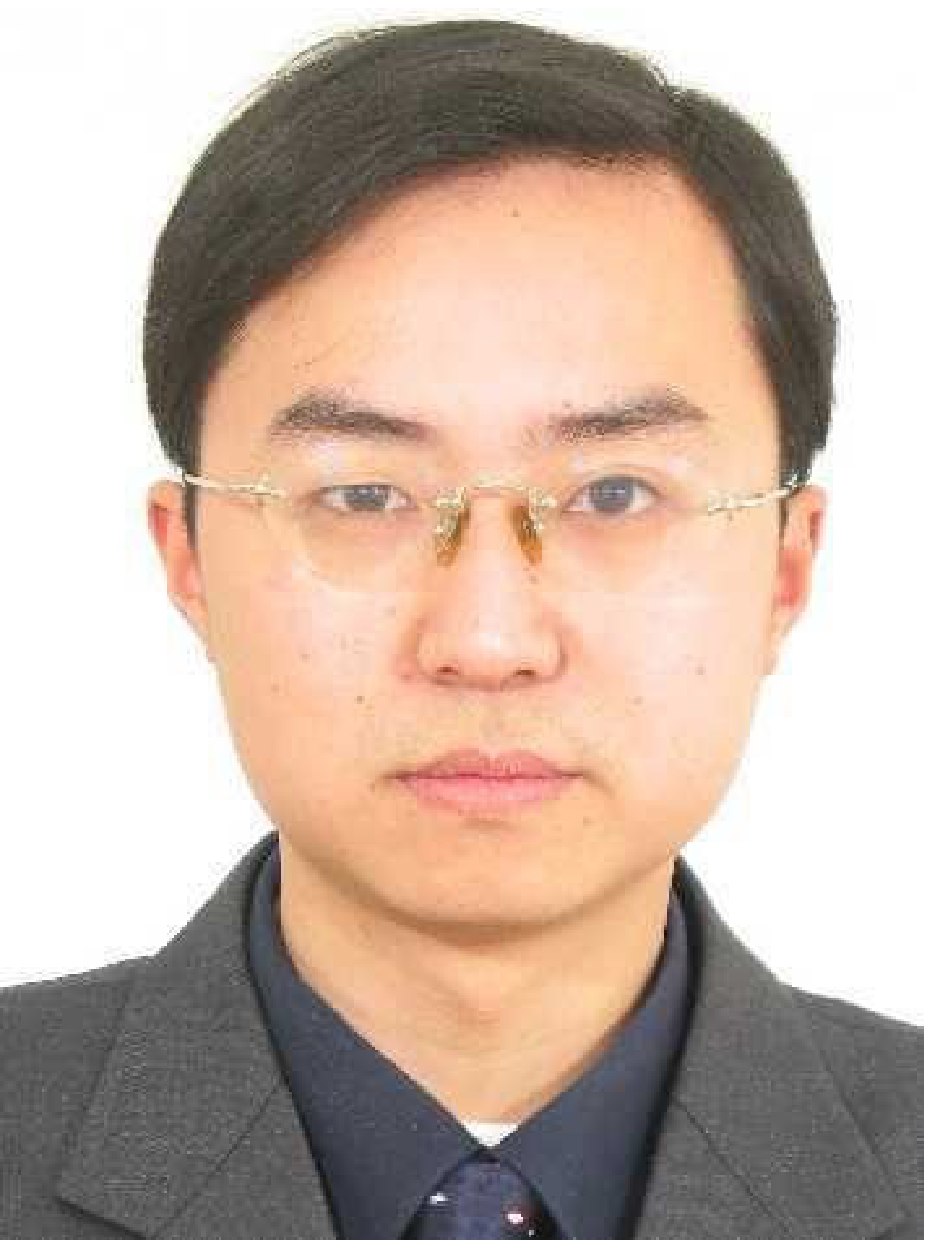}}]{Xiaohu Ge}
(M'09-SM'11) is currently a full Professor with the School of Electronic Information and Communications at Huazhong University of
 Science and Technology (HUST), China. He is an adjunct professor with the Faculty of Engineering and Information Technology at
 University of Technology Sydney (UTS), Australia. He received his PhD degree in Communication and Information Engineering from
 HUST in 2003. He has worked at HUST since Nov. 2005. Prior to that, he worked as a researcher at Ajou University (Korea) and
 Politecnico Di Torino (Italy) from Jan. 2004 to Oct. 2005. He was a visiting researcher at Heriot-Watt University, Edinburgh,
 UK from June to August 2010. His research interests are in the area of mobile communications, traffic modeling in wireless networks,
 green communications, and interference modeling in wireless communications. He has published more than 100 papers in refereed journals
  and conference proceedings and has been granted about 15 patents in China. He received the Best Paper Awards from IEEE Globecom 2010.
  He is leading several projects funded by NSFC, China MOST, and industries. He is taking part in several international joint projects,
  such as the EU FP7-PEOPLE-IRSES: project acronym WiNDOW (grant no. 318992) and project acronym CROWN (grant no. 610524).

Dr. Ge is a Senior Member of the China Institute of Communications and a member of the National Natural Science Foundation of China and
the Chinese Ministry of Science and Technology Peer Review College. He has been actively involved in organizing more than ten international
conferences since 2005. He served as the general Chair for the 2015 IEEE International Conference on Green Computing and Communications
(IEEE GreenCom). He serves as an Associate Editor for the IEEE ACCESS, Wireless Communications and Mobile Computing Journal (Wiley) and the
 International Journal of Communication Systems (Wiley), etc. Moreover, he served as the guest editor for IEEE Communications Magazine Special
  Issue on 5G Wireless Communication Systems.
\end{IEEEbiography}

\begin{IEEEbiography}[{\includegraphics[width=1in,height=1.25in,clip,keepaspectratio]{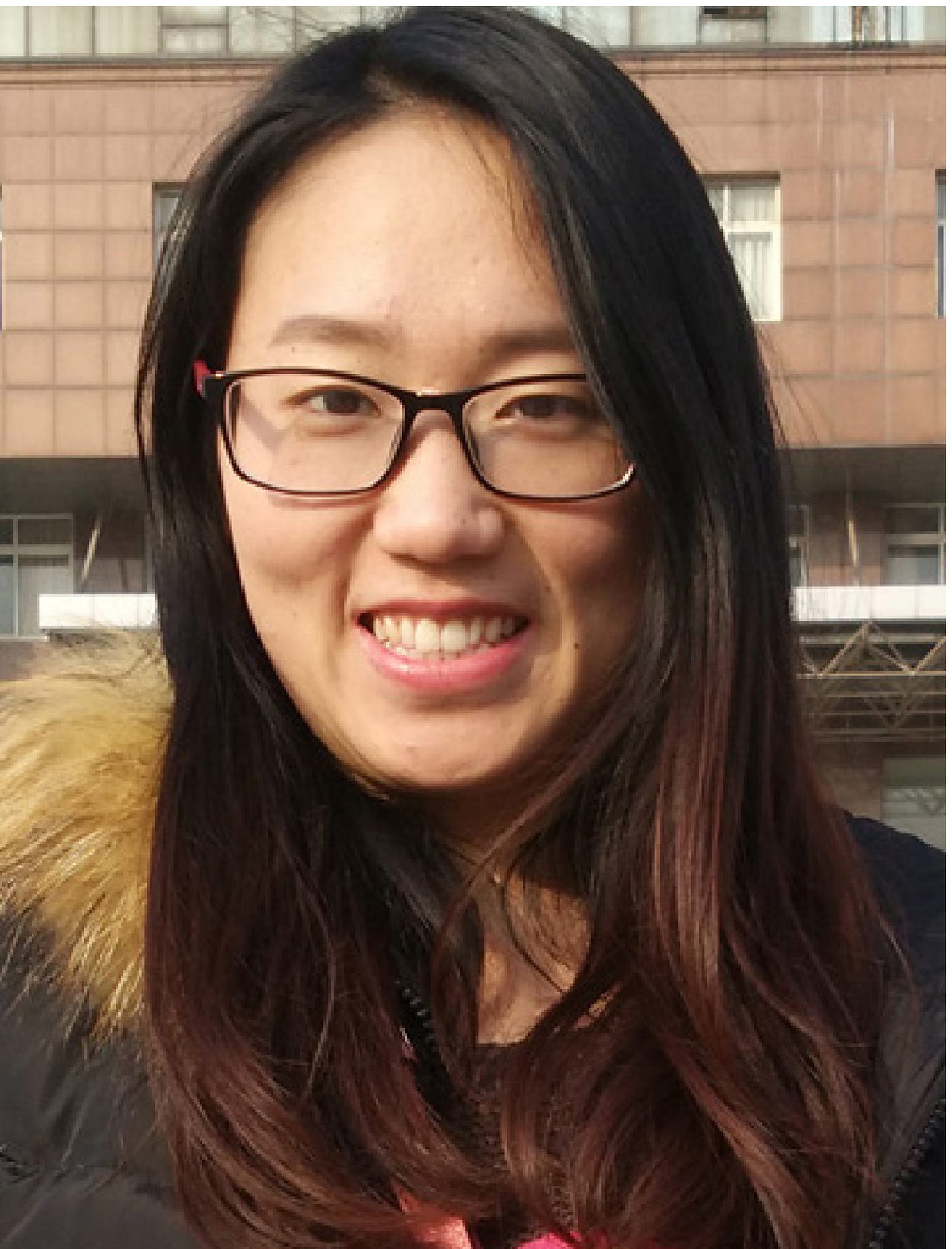}}]{Hui Cheng}
received her Bachelor¡¯s degrees in information engineering from Wuhan University of Technology,
 China, in 2013. She is currently working toward her Master¡¯s degree in communication and information systems at HUST.
She was an exchange student at the University of Linkoping, Sweden from June to September 2015. Her research
interests are in the fields of mobile backhaul traffic and user mobility models for small cell networks.
\end{IEEEbiography}

\begin{IEEEbiography}[{\includegraphics[width=1in,height=1.25in,clip,keepaspectratio]{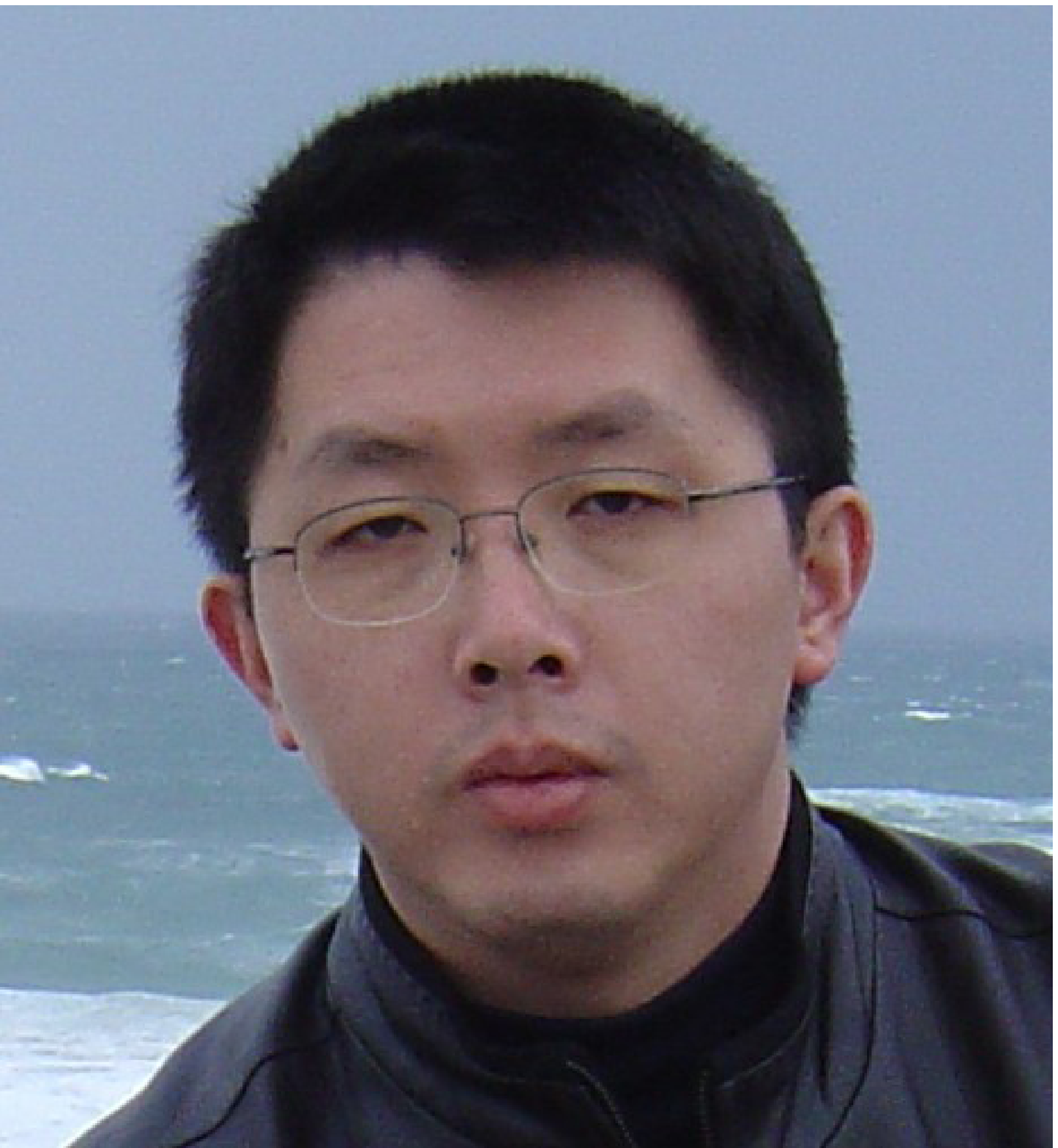}}]{Guoqiang Mao}
(S'98-M'02-SM'08) (e-mail:g.mao@ieee.org) received PhD in telecommunications engineering in 2002 from Edith Cowan University.He is with the
School of Computing and Communications, The University of Technology Sydney. He is also with Beijing University of Posts and Telecommunications
 and Huazhong University of Science and Technology. He was with the School of
Electrical and Information Engineering, the University of Sydney between 2002 and 2014. He joined the University of Technology Sydney
in February 2014 as Professor of Wireless Networking and Director of Center for Real-time Information Networks. The Center is among the
largest university research centers in Australia in the field of wireless communications and networking. He has published more than 150
papers in international conferences and journals, which have been cited more than 3500 times.  He is an editor of the IEEE Transactions
on Wireless Communications (since 2014), IEEE Transactions on Vehicular Technology (since 2010) and received ¡°Top Editor¡± award for
outstanding contributions to the IEEE Transactions on Vehicular Technology in 2011 and 2014. He is a co-chair of IEEE Intelligent Transport
Systems Society Technical Committee on Communication Networks. He has served as a chair, co-chair and TPC member in a large number of
international conferences. His research interest includes intelligent transport systems, applied graph theory and its applications in
telecommunications, wireless sensor networks, wireless localization techniques and network performance analysis.
\end{IEEEbiography}

\begin{IEEEbiography}[{\includegraphics[width=1in,height=1.25in,clip,keepaspectratio]{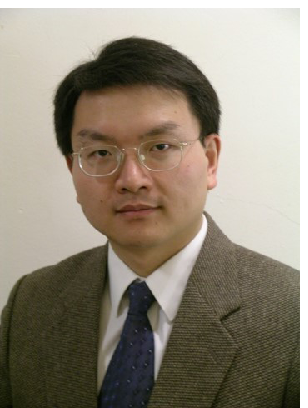}}]{Yang Yang}
(S'99-M'02-SM'10) received the BEng and MEng degrees in Radio Engineering from Southeast University, Nanjing, P. R. China, in 1996
 and 1999, respectively; and the PhD degree in Information Engineering from The Chinese University of Hong Kong in 2002.
Dr. Yang Yang is currently a professor with Shanghai Institute of Microsystem and Information Technology (SIMIT), Chinese Academy of Sciences,
 serving as the Director of CAS Key Laboratory of Wireless Sensor Network and Communication, and the Director of Shanghai Research Center for
Wireless Communications (WiCO). He is also an adjunct professor with the School of Information Science and Technology, ShanghaiTech University.
 Prior to that, he has served the Department of Electronic and Electrical Engineering at University College London (UCL), United Kingdom,
 as a Senior Lecturer; the Department of Electronic and Computer Engineering at Brunel University, United Kingdom, as a Lecturer; and the
 Department of Information Engineering at The Chinese University of Hong Kong as an Assistant Professor. His research interests include
 wireless ad hoc and sensor networks, software defined wireless networks, 5G mobile systems, intelligent transport systems, wireless testbed
  development and practical experiments.

Dr. Yang Yang has co-edited a book on heterogeneous cellular networks (2013, Cambridge University Press) and co-authored more than 100
technical papers. He has been serving in the organization teams of about 50 international conferences, e.g. a co-chair of Ad-hoc and Sensor
Networking Symposium at IEEE ICC¡¯15, a co-chair of Communication and Information System Security Symposium at IEEE Globecom¡¯15.
\end{IEEEbiography}

\begin{IEEEbiography}[{\includegraphics[width=1in,height=1.25in,clip,keepaspectratio]{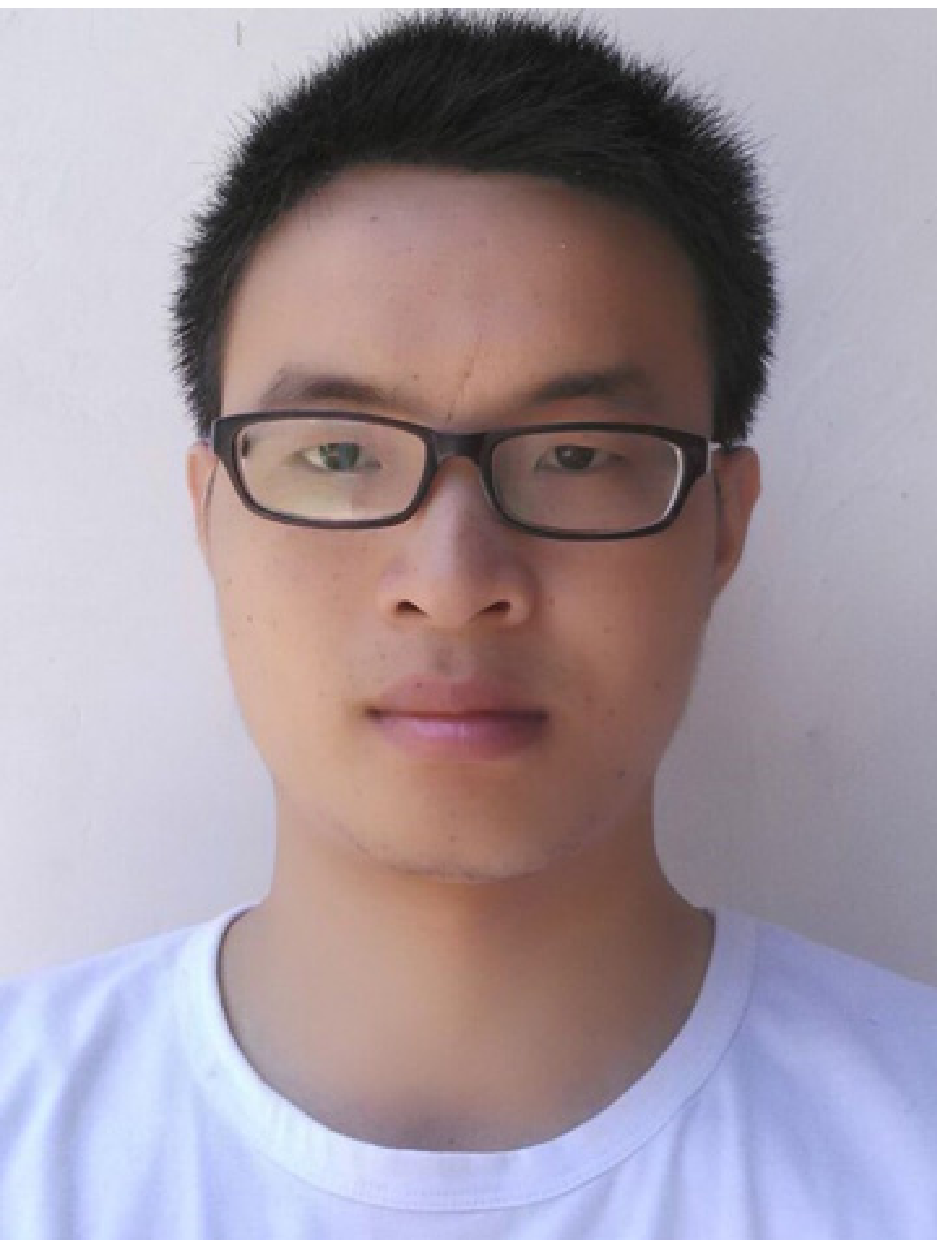}}]{Song Tu}
(songtu@mail.hust.edu.cn)received his B.E.degrees from Huazhong University of Science and Technology, China, in 2014. Now he continues to
study for a master¡¯s degree in the School of Electronic Information and Communications at the same university. His research interests are in
the area of green communications and distributed wireless networks.
\end{IEEEbiography}


\begin{thebibliography}{1}

\bibitem{ChenZhao14}
S. Chen and J. Zhao, ``The requirements, challenges and technologies for 5G of terrestrial mobile telecommunication," {\em IEEE Commun. Mag.}, vol. 52, no. 5, pp. 36--43, May. 2014.

\bibitem{ChenZhang15}
M. Chen, Y. Zhang, Y. Li, S. Mao, V. Leung, ``EMC: emotion-aware mobile cloud computing in 5G," {\em IEEE Network}, vol. 29, no. 2, pp. 32--38, Mar. 2015.

\bibitem{Ge16}
X. Ge, S. Tu, G. Mao, C.-X. Wang and T. Han, ``5G ultra-dense cellular networks," {\em IEEE Wireless Commun.}, vol. 23, no. 1, pp. 72--79, Feb. 2016.

\bibitem{Maviel12}
L. Maviel, Y. Lostanlen, J. Gorce, ``A hybrid propagation model for large-scale variations caused by vehicular traffic in small cells," {\em in Proc. IEEE GLOBECOM}, pp. 5021--5026, Dec. 2012.

\bibitem{Ge14}
X. Ge, H. Cheng, M. Guizani, T. Han, ``5G wireless backhaul networks: challenges and research advances," {\em IEEE Network}, vol. 28, no. 6, pp. 6--11, Nov. 2014.

\bibitem{Cheng14}
X. Cheng, C.-X. Wang, B. Ai, and H. Aggoune, ``Envelope level crossing rate and average fade duration of non-isotropic vehicle-to-vehicle Ricean fading channels," {\em IEEE Trans. Intell. Transp. Syst.}, vol. 15, no. 1, pp. 62--72, Feb. 2014.

\bibitem{Wang07}
C.-X. Wang, M. Patzold, and Q. Yao, ``Stochastic modeling and simulation of frequency correlated wideband fading channels," {\em IEEE Trans. Veh. Technol.}, vol. 56, no. 3, pp. 1050--1063, May. 2007.

\bibitem{ZhuBao12}
Y. Zhu, Y. Bao, B. Li, ``On maximizing delay-constrained coverage of urban vehicular networks," {\em IEEE J. Sel. Areas in Commun.}, vol. 30, no. 4, pp. 804--817, May. 2012.

\bibitem{Khabbaz12}
 M. J. Khabbaz, W. F. Fawaz, C. M. Assi, ``A probabilistic and traffic-aware bundle release scheme for vehicular intermittently connected networks," {\em IEEE Trans. Commun.}, vol. 60, no. 11, pp. 3396--3406, Aug. 2012.

 \bibitem{ChenHao15}
 M. Chen, Y. Hao, Y. Li, C. Lai, D. Wu, ``On the computation offloading at ad hoc cloudlet: architecture and service models," {\em IEEE Commun.}, vol. 53, no. 6, pp. 18--24, Jun. 2015.

\bibitem {LiZhangHaenggi15}
 C. Li, J. Zhang,  M. Haenggi, K. B. Letaief, ``User-centric intercell interference nulling for downlink small cell networks," {\em IEEE Trans. Commun.}, vol. 63, no. 4, pp. 1419--1431, Feb. 2015.

\bibitem{YuWang14}
J. Yu, Y. Wang, X. Lin, Q. Zhang, ``Statistical analysis of capacity in joint processing coordinated multi-point systems," {\em in Proc. IEEE/CIC ICCC}, pp. 496--501, Oct. 2014.

\bibitem{Sepulcre11}
 M. Sepulcre, J. Mittag, P. Santi, H. Hartenstein, ``Congestion and awareness control in cooperative vehicular systems," {\em Proc. IEEE JPROC}, vol. 99, no. 7, pp. 1260--1279, Jun. 2011.

\bibitem{Shafiee11}
K. Shafiee, A. Attar, V. C. M. Leung, ``Optimal distributed vertical handoff strategies in vehicular heterogeneous networks," {\em IEEE J. Sel. Areas in Commun.}, vol.29, no. 3, pp. 534--544, Mar. 2011.

\bibitem{Bernal-Mor15}
E. Bernal-Mor, V. Pla, J. Martinez-Bauset, L. Guijarro, ``Performance analysis of two-tier wireless networks with dynamic traffic, backhaul constraints and terminal mobility," {\em IEEE Trans. Veh. Technol.}, DOI:10.1109/TVT.2015.2397317.

\bibitem{LiXu14}
L. Li, Y. Xu, Lin Ma, ``Vertical handoff strategy on achieving throughput in vehicular heterogeneous network," {\em in Proc. IEEE VTC--Spring},  pp.1--5, May. 2014.

\bibitem{LiZhangXu14}
X. Li, H. Zhang, Q. Xu, ``Optimal access scheme for mobile vehicular small cells in layered heterogenous networks," {\em in Proc. IEEE IC-NIDC}, pp. 46--50, Sept. 2014.

\bibitem{Feteiha14}
M. F. Feteiha,  M. H. Qutqut, H. S. Hassanein,  ``Outage probability analysis of mobile small cells over LTE-A networks," {\em  IEEE IWCMC}, pp. 1045--1050, Aug. 2014.

\bibitem{Gesbert10}
 D. Gesbert, S. Hanly, H. Huang, S. Shamai Shitz, ``Multi-cell mimo cooperative networks: a new look at interference," {\em IEEE J. Sel. Areas in Commun.}, vol. 28, no. 9, pp. 1380--1408, Oct. 2010.

\bibitem{Baccelli14}
F. Baccelli,  A. Giovanidis, ``A stochastic geometry framework for analyzing pairwise-cooperative cellular networks," {\em IEEE Trans. Wireless Commun.}, vol. 14, no. 2, pp. 794--808, Sept. 2014.

\bibitem{Tanbourgi14}
R. Tanbourgi, S. Singh, J. G. Andrews, F. K. Jondral,  ``A tractable model for noncoherent joint-transmission base station cooperation," {\em IEEE Trans. Wireless Commun.}, vol. 13, no. 9, pp. 4959--4973, July. 2014.

\bibitem{Sakr14}
A. H. Sakr, E. Hossain,  ``Location-aware cross-tier coordinated multipoint transmission in two-tier cellular networks," {\em IEEE Trans. Wireless Commun.}, vol. 13, no. 11, pp. 6311--6325,  Aug. 2014.

\bibitem{ZhangYang12}
Q. Zhang, C. Yang, ``Transmission mode selection for downlink coordinated multipoint systems," {\em IEEE Trans. Veh. Technol.}, vol. 62, no. 1, pp. 465--471, Sept. 2012.

\bibitem{LiHu13}
Q. Li,  R. Q. Hu, Y. Qian, G. Wu, ``Intracell cooperation and resource allocation in a heterogeneous network with relays," {\em IEEE Trans. Veh. Technol.}, vol. 62, no. 4, pp. 1770--1784, May. 2013.

\bibitem{Agarwal14}
S. Agarwal, S. De, S. Kumar, Gupta, H.M. ``Qos-aware downlink cooperation for cell-edge and handoff users," {\em IEEE Trans. Veh. Technol.}, vol. 64, no. 6, pp. 2512--2527, Aug. 2014.

\bibitem{LiuNatarajan15}
C. Liu, B. Natarajan, H. Xia,  ``Small cell base station sleep strategies for energy efficiency," {\em IEEE Trans. Veh. Technol.}, DOI:10.1109/TVT.2015.2413382

\bibitem{Song14}
H. Song, X. M. Fang, L. Yan, ``Handover scheme for 5G CU plane split heterogeneous network in high-speed railway," {\em IEEE Trans. Veh. Technol.}, vol. 63, no. 9, pp. 4633--4646, Apr. 2014.

\bibitem{Ferenc07}
J. S. Ferenc and Z. Neda, ``On the size distribution of poisson voronoi cells," {\em Physica A}, vol. 385, no. 2, pp. 518--526, Nov. 2007.

\bibitem{Stoyan96}
D. Stoyan, W. S. Kendall, and J. Mecke, {\em Stochastic Geometry and Its Applications}, 2nd edition. Wiley, 1996.

\bibitem{Foss96}
S. G. Foss and S. A. Zuyev, ``On a voronoi aggregative process related to a bivariate Poisson process,"  {\em Adv Appl Probab. }, vol. 28, no. 4, pp. 965--981, Dec. 1996.

\bibitem{Haenggi05}
M. Haenggi, ``On distances in uniformly random networks," {\em IEEE Trans. Inf. Theory}, vol. 51, no. 10, pp. 3584--3586, Oct. 2005.

\bibitem{Wang09}
C. Wang, S. J. Tang, X. Y. Li, C. J. Jiang, ``Multicast capacity of multihop cognitive networks,"  {\em IEEE MOBHOC}, pp. 274--283, Oct. 2009.

\bibitem{Fan15}
 C. Fan, Y. J. Zhang, X. Yuan, ``Dynamic nested clustering for parallel phy-layer processing in Cloud-RANs," {\em  IEEE Trans. Wireless Commun.}, DOI:10.1109/TWC.2015.2496953.

\bibitem{Dighe03}
P. A. Dighe, R. K. Mallik, and S. S. Jamuar, ``Analysis of transmitreceive diversity in rayleigh fading," {\em IEEE Trans. Commun.}, vol. 51, no. 4, pp. 694--703, Apr. 2003.

\bibitem{Ge11}
X. Ge, K. Huang, C. Wang, X. Hong, X. Yang, ``Capacity analysis of a multi-cell multi-antenna cooperative cellular network with co-channel interference," {\em IEEE Trans. Wireless Commun.}. vol. 10, no. 10, pp. 3298 -- 3309, Oct. 2011.

\bibitem{Haenggi09}
M. Haenggi, G. J. Andrews, F. Baccelli, O. Dousse, and M. Franceschetti, ``Stochastic geometry and random graphs for the analysis and design of wireless networks," {\em IEEE J. Sel. Areas in Commun.}. vol. 27, no. 7, pp. 1029--1046, Sept. 2009.

\bibitem{Nigam14}
G. Nigam,  P. Minero, and M. Haenggi, ``Coordinated multipoint joint transmission in heterogeneous networks," {\em IEEE Trans. Commun.}, vol. 62, no. 11, pp. 4134--4146, Nov. 2014.

\bibitem{LiZhangLetaief14}
C. Li, J. Zhang, K. B. Letaief, ``Throughput and energy efficiency analysis of small cell networks with multi-antenna base stations," {\em IEEE Trans. Wireless Commun.}, vol. 13, no. 5, pp. 2505--2517, Mar. 2014.

\bibitem{Robson12}
J. Robson, ``Small cell backhaul requirements," {\em NGMN White Paper}, pp. 1--40, Jun. 2012.

\bibitem{Widjaja09}
I. Widjaja,  H. La Roche, ``Sizing X2 bandwidth for inter-connected eNBs," {\em in Proc. IEEE VTC--Fall}, pp. 1--5, Sept. 2009.

\end{thebibliography}
\end{document}